\documentclass[12pt,preprint]{emulateapj}

\shorttitle{Modeling Star-Forming Galaxies}
\shortauthors{Levesque et al.}
\slugcomment{DRAFT \today}

\begin{document}

\title{Theoretical Modeling of Star-Forming Galaxies I. Emission Line Diagnostic Grids for Local and Low-Metallicity Galaxies}

\author{Emily M. Levesque$^{1}$, Lisa J. Kewley, Kirsten L. Larson}
\affil{Institute for Astronomy, University of Hawaii, 2680 Woodlawn Dr., Honolulu, HI 96822}
\email{emsque@ifa.hawaii.edu}

\begin{abstract}
We use the newest generation of the Starburst99/Mappings code to generate an extensive suite of models to facilitate detailed studies of star-forming galaxies and their ISM properties, particularly at low metallicities. The new models used include a rigorous treatment of metal opacities in the population synthesis modeling and more detailed dust physics in the photoionization code. These models span a wide range of physical parameters 
including metallicity, ionization parameter, and the adoption of both a instantaneous burst and continuous star formation history. We examine the agreement between our models and local ($z < 0.1$) star-forming galaxy populations from several large datasets, including the Sloan Digital Sky Survey, the Nearby Field Galaxy Survey, and samples of blue compact galaxies and metal-poor galaxies. We find that models adopting a continuous star formation history reproduce the metallicity-sensitive line ratios observed in the local population of star-forming galaxies, including the low-metallicity sample. However, we find that the current codes generate an insufficiently hard ionizing radiation field, leading to deficiencies in the [SII] fluxes produced by the models. We consider the advantages and short-comings of this suite of models, and discuss future work and improvements that can be applied to the modeling of star-forming galaxies.
\end{abstract}

\footnotetext[1]{Predoctoral Fellow, Smithsonian Astrophysical Observatory, 60 Garden St., MS-20, Cambridge, MA 02138}

\section{Introduction}
\label{Sec-intro}
Robust analysis of star-forming galaxy emission-line spectra can provide constraints on key physical parameters of the ionizing radiation field and the interstellar medium (ISM). The star formation rates (SFRs) of these galaxies can be estimated from luminosities of the H$\alpha$ line (Hunter \& Gallagher 1986, Kennicutt 1998, Bicker \& Fritze-v.\ Alvensleben 2005, Kewley et al.\ 2007) or the [OII] line (Gallagher et al.\ 1989, Kennicutt 1998, Rosa-Gonz\'{a}lez et al.\ 2002, Kewley et al.\ 2004, Moustakas et al.\ 2006), and multiple studies have examined the use of the H$\beta$ equivalent width to estimate the age of the young stellar population in galaxies (Schaerer \& Vacca 1998, Gonzalez Delgado et al.\ 1999, Fernandes et al.\ 2003, Martin-Manjon et al.\ 2008). Metallicity is another critical parameter in the study of star-forming galaxy ISM environments, shedding light on star formation histories and subsequent chemical evolution. A variety of optical emission line ratio diagnostics have been presented and employed to determine metallicities, including measuring the [OIII] line ratios to determine electron temperatures, and therefore abundances (e.g. Peimbert 1967, Garnett 1992), and use of the [NII]$\lambda$6584/H$\alpha$ and [NII]$\lambda$6584/[OIII]$\lambda$5007 ratio diagnostics from Pettini \& Pagel (2004), the [NII]$\lambda$6584/[OII]$\lambda$3727 ratio from Kewley \& Dopita (2002), and the ([OIII]$\lambda$5007 + [OIII]$\lambda$4959 + [OII]$\lambda$3727)/H$\beta$ (R$_{23}$) ratio (e.g., Pagel et al.\ 1979, McGaugh 1991, Zaritsky et al.\ 1994, Pilyugin 2000, Charlot \& Longhetti 2001, Kewley \& Dopita 2002, Kobulnicky \& Kewley 2004).

Baldwin et al.\ (1981) present the technique of plotting optical emission line ratios, such as [NII]$\lambda$6584/H$\alpha$ vs. [OIII]$\lambda$5007/H$\beta$, on a series of diagnostic diagrams to separate extragalactic objects according to their primary excitation mechanisms. These and other diagrams were later used by Veilleux \& Osterbrock (1987) to derive a semi-empirical classification scheme to distinguish between star-forming galaxies and active galactic nuclei. Kewley et al.\ (2001) used these same diagrams to derive a purely theoretically classification scheme, which was later extended by Kauffmann et al.\ (2003) and Stasinska et al.\ (2006) using data from the Sloan Digital Sky Survey. Emission line diagnostic diagrams can also be used to probe the shape of the far-ultraviolet ionizing spectrum of a galaxy (Dopita et al.\ 2000, Kewley et al.\ 2001, Kewley \& Dopita 2002). As a result, diagnostic ratio grids are often employed to test the agreement of stellar population synthesis and photoionization model grids with emission-line ratios measured in observations of star-forming galaxies. These comparisons effectively illustrate the evolution and improvements of such models, and are also useful in highlighting the shortcomings of different grids and the challenges faced in modeling these galaxies and their emission spectra.

Kewley et al.\ (2001) use both the Starburst99 (Leitherer et al.\ 1999) and Pegase (Fioc \& Rocca-Volmerage 1997) evolutionary synthesis models in conjunction with the Mappings III photoionization code (Binette et al.\ 1985, Sutherland \& Dopita 1993) to compute photoionization models that they compare to a sample of 157 warm IR starburst galaxies on a variety of optical emission line ratio diagnostic diagrams. They find that assuming a continuous star formation history is more realistic than the assumption of a single instantaneous burst. Moy et al.\ (2001) find a similar result; using the Pegase evolutionary synthesis models and the CLOUDY photoionization code (Ferland 1996) they model young stellar populations along with underlying stellar populations, and find that underlying populations with continuous star formation histories are more compatible with observed emission line ratios in starburst galaxies. Fernandes et al.\ (2003) also find support for modeling underlying stellar populations in their population synthesis analysis of starburst and HII galaxies, using the Bruzual \& Charlot (1993) evolutionary synthesis code.

Kewley et al.\ (2001) do, however, find that the ionizing spectra produced by the Starburst99 models are not hard enough in the far ultraviolet (FUV) region of the spectrum to reproduce the observed line ratios, and propose that including the effects of continuum metal opacities in stellar atmospheres should be a way of improving the models' accuracy. This deficiency in the FUV ionizing spectrum is noted in Panuzzo et al.\ (2003) and Magris et al.\ (2003). Starburst99 models at the time used older stellar atmosphere models that do not include treatments of opacities beyond hydrogen and helium (Schmutz et al.\ 1992), or else are not complete at the high temperatures or masses that prove critical when modeling the FUV ionizing radiation field (Rauch 1997, Lejeune et al.\ 1997, 1998).

Another challenge that is common among previous stellar population synthesis and photoionization models is a difficulty in modeling the environments of metal-poor galaxies (e.g. Fernandes et al.\ 2003, Dopita et al.\ 2006, Martin-Manjon et al.\ 2008). Dopita et al.\ (2006) use the latest Starburst99 code from V\'{a}zquez \& Leitherer (2005) and the Mappings III photoionization code to generate models of isobaric dusty HII regions, as well as integrated galaxy spectra that model the galaxies as composites of multiple HII regions. Dopita et al.\ (2006) removes the ionization parameter as a free parameter, fixing the ISM pressure and central cluster mass and instead allowing ionization parameter to vary with time, pointing out that ionization parameter also varies with metallicity. Dopita et al.\ (2006) find that these models still do not reproduce the observed emission line diagnostic ratios of lower-metallicity galaxies ($Z < 0.4Z_{\odot}$). They also note that the ionizing fluxes are not hard enough to agree with the integrated spectra.  Martin-Manjon et al.\ (2008, 2009) take a different approach, using sets of stellar yields from Gavilan et al.\ (2005) to model the chemical evolution of HII galaxies, allowing metallicity to evolve with age rather than generating a grid with a range of fixed metallicities. They also use the newer generation of CLOUDY (Ferland et al.\ 1998) to model environments that have undergone multiple bursts of star formation (Martin-Manjon et al.\ 2009), but still find that these models don't account for the most metal-deficient HII galaxies in their sample.

Metal-poor galaxies are an important avenue of study in their own right. With a gas-phase oxygen abundance of log(O/H) + 12 $<$ 7.9 (Yin et al.\ 2007), they offer a relatively pristine pre-enchriment ISM in which star-formation, as well as current and previous episodes of enrichment, can be examined (Brown et al.\ 2008). These galaxies also pose challenges to scenarios seeking to reproduce their metal-poor environments through a variety of evolutionary mechanisms (see Kewley et al.\ 2007 and references therein). It has also recently been proposed that the host galaxies of long-duration gamma-ray bursts belong to the metal-poor galaxy population (Stanek et al.\ 2006, Fruchter et al.\ 2006, Kewley et al.\ 2007, Modjaz et al.\ 2008). Proper modeling of these galaxies is critical to our understanding of the star formation processes and evolving stellar populations present in such environments.

Here we present models generated using the V\'{a}zquez \& Leitherer (2005) Starburst99 stellar population synthesis code and the latest generation of the Mappings III code, with recent improvements that include a more rigorous treatment of dust (Groves et al.\ 2004). Our models are tailored towards addressing the difficulties found in past work with producing a harder FUV spectrum; we do this by adopting the WMBASIC stellar atmosphere models of Pauldrach et al.\ (2001) and the CMFGEN Hillier \& Miller (1998) atmospheres, both of which include the rigorous treatments of continuum metal opacities that were suggested in Kewley et al.\ (2001). Our grid was also designed to precisely model the emission line flux of the local galaxy population, including low-metallicity galaxies, adopting the full range of metallicities available in the evolutionary tracks published by the Geneva group (Schaller et al.\ 1992; Schaerer et al.\ 1993a, 1993b; Charbonnel et al.\ 1993). 

In this paper we describe our new grid of stellar population synthesis models, detailing the inputs and free parameters that we adopt when generating the grid and examining the ionizing FUV spectra that are produced by Starburst99 in detail (\S~2). We consider how these spectra affect the behavior of a variety of optical emission line diagnostics (\S~3). With these diagnostic ratios, we generate a series of optical emission line diagnostic diagrams, and compare our model grids to spectra of a variety of nearby ($z \le 0.1$) galaxy populations to assess their agreement (\S~4). Finally, we consider the results of these comparisons and discuss potential future work in this area (\S~5).

\section{Starburst99/Mappings III Model Grids}
\label{Sec-mod}
\subsection{Model Grid Parameters}
\label{Sec-params}
To model our sample of galaxies we have used the Starburst99 code (Leitherer
et al.\ 1999, V\'{a}zquez \& Leitherer 2005) to generate theoretical spectral energy distributions
(SEDs), which in turn were used in the Mappings III photoionization models to produce model
galaxy spectra that could be compared to our observations.

Starburst99 is an evolutionary synthesis code that can be used to generate synthetic ionizing far-ultraviolet (FUV) radiation spectra as a function of metallicity, star formation history, and the age and evolution of the stellar populations. These populations are produced by use of model stellar atmospheres and spectra along with evolutionary tracks for massive stars. For this work, we have used a Salpeter initial mass function (Salpeter 1955) with a 100M$_{\odot}$ upper mass boundary, along with Starburst99's Pauldrach/Hillier model atmospheres. These employ the WMBASIC wind models of Pauldrach et al.\ (2001) for younger ages when O stars dominate the luminosity ($<$ 3 Myr), and the CMFGEN Hillier \& Miller (1998) atmospheres for later ages at which Wolf-Rayet stars are dominant. This differs from the Starburst99 models presented in Dopita et al.\ (2000) and Kewley et al.\ (2001), which adopt the plane-parallel Lejeune et al.\ (1997) grid of atmosphere models along with the Schmutz et al.\ (1992) extended model atmospheres for stars with higher winds. The Schmutz et al.\ (1992) models, which include the critical Wolf-Rayet phase, only include continuous opacities for hydrogen and helium, neglecting what are expected to be considerable effects from continuum metal opacities. Kewley et al.\ (2001) suggest this as a potential shortcoming in their models, proposing that the inclusion of continuum metal opacities will result in a fraction of higher-energy radiation being absorbed and reemitted at lower energies, in the region of the FUV spectrum that is responsibly for ionizing the optical emission lines. The Pauldrach et al.\ (2001) and Hillier \& Miller (1998) model atmospheres address this shortcoming by including rigorous non-LTE treatments of metal opacities. In conjunction with the Pauldrach/Hillier atmospheres we also adopt two different sets of evolutionary tracks produced by the Geneva group, and consider the particulars of these tracks' mass loss rates (see Section \S~2.2). Starburst99 generates the final synthetic FUV spectrum output through use of the isochrone synthesis method first introduced by Charlot \& Bruzual (1991), fitting isochrones to the evolutionary tracks across different masses rather than discretely assigning stellar mass bins to specific tracks. 

The resulting FUV spectrum is then taken as input by the Mappings III code. The Mappings shock and photoionzation code was originally developed by Binette et al.\ (1985), improved by Sutherland \& Dopita (1993), and used in Dopita et al.\ (2000) and Kewley et al.\ (2001). The recent improvements to the Mappings code, used in Dopita et al. (2006) and Snijders et al. (2007), include a more sophisticated treatment of dust (Groves et al.\ 2004) that is adopted in our models, which properly treats the effects of absorption, charging, and photoelectric heating by the grains - for a more detailed discussion, see Groves et al.\ (2004) and Snijders et al.\ (2007). Taking the synthetic ionizing FUV spectrum output of Starburst99 and an adopted nebular geometry model, which we assume to be plane-parallel, we select a variety of electron densities and ionization parameters as inputs. Using these parameters in Mappings III we computed a complete grid of plane-parallel isobaric photoionization models.

When generating our Starburst99/Mappings III model spectra, we adopted a broad grid of input parameters to facilitate comparison with a wide range of galaxy samples:

{\it Star Formation History (SFH)}: We model both a zero-age instantaneous burst of star formation, with a fixed mass of $10^6 M_{\odot}$, and a continuous SFH where the star formation rate (SFR) is constant at a rate of 1 $M_{\odot}$ per year, starting from an initial time and assuming a stellar population that is large enough to fully sample the IMF at high masses.

{\it Metallicity}: We adopt the full range of metallicities available from the evolutionary tracks of the Geneva group, which includes five metallicities of $z = 0.001$ ($Z = 0.05Z_{\odot}$), $z = 0.004$ ($Z = 0.2Z_{\odot}$), $z = 0.008$ ($Z = 0.4Z_{\odot}$), $z = 0.02$ ($Z = Z_{\odot}$), and $z = 0.04$ ($Z = 2Z_{\odot}$).

{\it Evolutionary Tracks}: We adopt the two evolutionary tracks of the Geneva group that are currently available in Starburst99: the Geneva ``Standard" mass loss tracks, and the Geneva ``High" mass loss tracks. The differences in these tracks are discussed in more detail in Section \S~2.2.

{\it Age}: We generate models ranging in age from 0 to 5 Myr in the case of an instantaneous burst star formation history in 0.5 Myr increments. For the continuous SFH models we adopt a constant age of 5 Myr, the age at which a continuous SFH stellar population reaches equilibrium (Kewley et al.\ 2001). Further discussion of the age range of these model grids can be found in \S~2.3.

{\it Ionization parameter}: The ionization parameter $q$ (cm s$^{-1}$) can be thought of as the maximum velocity possible for an ionization front being driven by the local radiation field. The value itself is calculated for the inner surface of the nebula. By this definition $q$ relates to the dimensionless ionization parameter $\mathcal{U}$ by $\mathcal{U} \equiv q/c$. Rigby \& Rieke (2004) find a range of $-3 <$ log $\mathcal{U} < -1.5$ for the dimensionless ionization parameter in local starburst galaxies. In our model grid, we adopted seven different values for our ionization parameter $q$, where $q = 1 \times 10^7, 2 \times 10^7, 4 \times 10^7, 8 \times 10^7, 1 \times 10^8, 2 \times 10^8,$ and $4 \times 10^8$ cm s$^{-1}$). These values correspond to dimensionless ionization parameters of log $\mathcal{U} \approx -3.5,-3.2,-2.9,-2.6,-2.5,-2.2,$ and $-1.9$, respectively. While these are slightly lower than the log $\mathcal{U}$ values found in Rigby \& Rieke (2004), they are similar to the range of ionization parameters adopted in Kewley (2001) and Snijders et al.\ (2007).

{\it Electron density}: Dopita et al.\ (2000) find $n_e$ = 10 cm$^{-3}$ to be typical of giant extragalactic HII regions, while Dopita et al.\ (2006) constrain the electron density in their models of HII regions to $n_e \lesssim 100$ cm$^{-3}$. In addition, Kewley et al.\ (2001) find an average electron density $n_e$ = 350 cm$^{-3}$ for the Kewley et al.\ (2000) sample of warm infrared starburst galaxies. Since we wish to compare our models to normal and low-metallicity star-forming galaxies, which are expected to have lower $n_e$ than those found in the gas-rich warm infrared galaxies, we adopt two different electron densities in our full model grid, adopting both $n_e = 10$ cm$^{-3}$ and $n_e = 100$ cm$^{-3}$. We assume an isobaric density structure for these models, and thus $n_e$ is specified by the dimensionless pressure/mean temperature ratio. We found that the lower $n_e = 10$ cm$^{-3}$ produced only a very slight decrease in flux for the more $q$-sensitive lines at higher metallicities. For the remainder of this paper, we present results that adopt $n_e = 100$ cm$^{-3}$, following the findings of Kewley et al.\ (2001) for starburst galaxies.

\subsection{Stellar Evolutionary Tracks}
\label{Sec-tracks}
Currently, Starburst99 includes two different evolutionary tracks produced by the Geneva
group, which differ primarily in their treatment of mass loss rates for massive stars. Mass loss rates are a critical parameter when considering the contributions of massive stars to ISM enrichment (Maeder \& Conti 1994).

The ``standard" (STD) mass loss evolutionary tracks were originally published in a series of papers by the Geneva group (Schaller et al.\ 1992; Schaerer et al.\ 1993a, 1993b; Charbonnel et al.\ 1993). These models adopt mass loss rates throughout the HR diagram from de Jager et al.\ (1988) that are scaled with metallicity according to the models of Kudritzki et al.\ (1989), where $\dot M \propto Z^{0.5}$. Wolf-Rayet (WR) stars are an exception - the mass loss rates for these stars are taken from Langer (1989) and Conti (1988) and include no correction for initial metallicity effects.

The ``high" mass loss evolutionary tracks (HIGH), published in Meynet et al.\ (1994), include enhanced mass loss rates, meant to more accurately reproduce observations of low-luminosity Wolf-Rayet stars and blue-to-red supergiant ratios in the Magellanic Clouds (Schaller et al.\ 1992, Meynet 1993). The adopted mass loss rates are derived by doubling the rates adopted by the ``standard" grid from de Jager et al.\ (1988), as well as doubling the ``standard" mass loss rate assumed for late-type WN-type WR stars. The mass-loss rates for early-type WN WR stars and later stages of WR stars (WC and WO) were left unchanged - for a complete discussion, see Meynet et al.\ (1994). Again, the mass loss rates of WR stars are left uncorrected for initial metallicity effects.

While many advances have since been made in our understanding of stellar physics, adopting these tracks in our stellar population synthesis models is still scientifically sound. The STD mass loss tracks are the more applicable of the two when considering recent work on the effect that wind clumping has on mass loss rates (Crowther et al.\ 2002); however, the HIGH mass loss tracks produce a reasonable approximation of the enhanced mass loss rates resulting from the effects of rotation, when surface mixing results in an earlier start of the WR phase (Meynet, private communication). Rotation is an important component of stellar evolution that is expected to have considerable influence on the Starburst99 ionizing spectrum and the agreement of these models with observations at low metallicity in particular (Leitherer 2008). Since the effects of rotation are not explicitly included in these tracks, the HIGH mass loss tracks can be considered a more appropriate approximation of the rotation-driven mass loss undergone by massive stars.

\subsection{Starburst99 Ionizing Spectra}
\label{Sec-FUV}
The far-ultraviolet (FUV) ionizing spectra produced by Starburst99 are primarily influenced by age and
metallicity. The effects of a changing model age derive largely from the evolution of the massive stellar population, and are easily examined in our models which adopt an instantaneous burst model of star formation - this star formation history allows us to observe the effects of a single stellar population that is formed at 0 Myr and evolves uniformly.

By contrast, for the continuous SFH models, we set the age constant at 5 Myr. This age describes an active (emission-line) star-forming galaxy, where the number of stars being formed is equal to the number of stars dying. At younger ages this equilibrium is not yet reached, and at older ages there is little to no evolution in the FUV ionizing spectrum produced by the stellar population. This is consistent with the evolution of continuous SFH models for starburst galaxies, as described in Kewley et al. (2001).

Figure 1 (left) shows the FUV spectra generated when adopting an instantaneous burst SFH and the HIGH evolutionary tracks, plotted from 0 Myr to 6 Myr in 1 Myr increments for our full range of metallicities. The hardness of the spectra decreases with age for this star formation history, most noticeably in the higher-energy regime of the spectra (100-300\AA), a result of the massive stellar population evolving out of the OB main sequence phase. 

It is apparent that the behavior of the FUV spectrum differs dramatically across the different metallicities. The low-metallicity spectra maintain a significantly harder ionizing spectrum throughout their evolution as compared to the $Z = Z_{\odot}$ and $Z = 2Z_{\odot}$ model spectra. Since high-metallicity stars spend a larger fraction of their high-energy photons ionizing their own atmospheric metals (due to the increased effects of line blanketing), there is a resulting depletion of high-energy photons available to ionize the surrounding ISM, which leads to a softer radiation field (Snijders et al.\ 2007). The effective temperatures of massive stars are also higher at lower metallicities across similar spectral types due to a shift of the evolutionary limits of hydrodynamic equilibrium (the Hayashi limit) to warmer temperatures at lower metallicities (Elias et al.\ 1985; Levesque et al.\ 2006), resulting in a hotter environment and harder spectrum in the case of our low-metallicity spectra. Finally, main sequence lifetimes are longer at low metallicities as a result of lower mass loss rates, leading to a greater amount of time spent in the hot main-sequence evolutionary phases and a larger contribution to the ionizing radiation field that extends to later ages (e.g. Meynet et al.\ 1994, Maeder \& Conti 1994).

Figure 1 (left) also indicates that the Wolf-Rayet (W-R) phase for these galaxies contributes to a hardening of the ionizing spectrum, producing a distinctive bump in the high-energy region of the spectrum. The net effect of this population is only visible in the high-Z ionizing spectra, appearing from 3 to 5 Myr in the $Z = Z_{\odot}$ spectrum, and from 4 to 6 Myr in the $Z = 2Z_{\odot}$. The wind-driven ISM enhancement by W-R stars is stronger and longer at high metallicities, and the minimum mass required to reach the W-R stage decreases at higher metallicities (taken to be 85 M$_{\odot}$ at $Z = 0.05Z_{\odot}$ as compared to 40M$_{\odot}$ at $Z = Z_{\odot}$ for the Geneva models used here; Schaller et al.\ 1992).

However, the most pronounced change in the spectra occurs at ages later than 5 Myr. For the 6 Myr model, there is a dramatic drop in the hardness of the spectra at $\sim$ 225 \AA, coinciding almost perfectly with the ionizing wavelength of [OIII]. This behavior is consistent with the origin of the FUV spectrum for this SFH. An instantaneous burst of star-formation at 0 Myr will result in a single coeval population of massive stars as the sole source of ionizing radiation, decreasing the hardness of the FUV spectrum at later ages as the hottest massive stars (and therefore the dominant contributors of ionizing radiation), evolve rapidly off the main sequence and end their lives. The lifetimes employed in the Geneva models give the explanation for the behavior of the FUV spectra shown here; the main-sequence (H-burning phase) lifetimes of a typical hot massive star of initial mass 40 M$_{\odot}$ all terminate at an age of 5 Myr or earlier (Schaller et al.\ 1992, Schaerer et al.\ 1993b, Charbonnel et al.\ 1993). As a result, beyond 5 Myr the dominant massive stellar population is comprised mainly of lower-mass ($\leq$ 25 M$_{\odot}$) stars that are beginning to evolve redwards on the Hertzsprung-Russell diagram,  and producing insufficient ionizing radiation at shorter wavelengths ($\lesssim$ 225 \AA) to generate the line fluxes we observe for highly ionized species. Because of this lack of ionizing photons, we restrict our photoionization grids generated for an instantaneous star formation history to 5 Myr and younger.

In practice, stellar populations in star-forming galaxies are thought to originate from episodic bursts of star formation, or from a continuous SFH. For example, Izotov \& Thuan (2004) find evidence of current ongoing star formation with an age of about 4 Myr in the extremely low-metallicity galaxy I Zw 18. In addition, they detect supergiant populations that indicate an intense star formation episode occuring 10-15 Myr ago and evidence of still older stellar populations with ages of hundreds of Myr. Noeske et al.\ (2007a, 2007b) examine 2905 star-forming galaxies from the All Wavelength Extended Groth Strip International Survey (AEGIS), and find that the dominant star formation mechanism in these galaxies appears to be a gradual decline of the average star formation rate, as opposed to a series of episodic bursts that decrease in frequency. Previous grids of stellar population synthesis models also find that the assumption of a continuous SFH, with the presence of underlying older stellar populations, is more appropriate than a single instantaneous burst in most cases (Kewley et al.\ 2001, Moy et al.\ 2001, Fernandes et al.\ 2003, Barton et al.\ 2003). This suggests that, rather than modeling a {\it single} burst of star-formation at 0 Myr (the current approach in Starburst99 for a burst-like SFH), a continuous SFH or multiple bursts of star formation through time would be a more realistic treatment for modeling spectra of star-forming galaxies (see also Lee et al.\ 2009). The instantaneous burst models presented here can therefore be employed mainly as an {\it approximation} of a burst-like SFH, representing a lower limit on star formation, while our continuous star formation models can be considered an upper limit.

After examining the behavior of the FUV spectra when adopting the HIGH evolutionary tracks, we can then compare these to the FUV spectra produced from the STD evolutionary tracks, subtracting the two spectra to examine their differences in greater detail (Figure 1, right). In this comparison, we can see that the largest differences occur at higher metallicities ($Z = 0.4Z_{\odot}, Z_{\odot}$, and 2$Z_{\odot}$) and in the 3 to 6 Myr age range, coinciding with the ionizing flux contribution of the mass-loss-sensitive Wolf-Rayet phase. Additional discrepancies at later ages focus on the age of the 225\AA\ drop, which is also expected to vary slightly with mass loss.

Finally, we find that the FUV spectra shown here are harder than those shown in Kewley et al.\ (2001) for both the PEGASE and the Starburst99 stellar population synthesis models. The increased hardness between 225\AA\ and 1000\AA\ is particularly notable, with log($\lambda$F$_{\lambda}$, ergs s$^{-1}$ M$_{\odot}^{-1}$) $\approx$ 25 to 30 for the solar-metallicity Kewley et al.\ (2001) models as compared to log($\lambda$F$_{\lambda}$, ergs s$^{-1}$ M$_{\odot}^{-1}$) $\approx$ 35 to 45 for the solar-metallicity models shown here. This corresponds to the 1 to 4 Ryd region that Kewley et al.\ (2001) cite as potentially benefiting from more rigorous treatments of continuum metal opacities, which we have adopted here via the inclusion of the Pauldrach et al.\ (2001) and Hillier \& Miller (1998) model atmospheres. These FUV spectra are therefore a notable improvement over the Kewley et al.\ (2001) grids, with higher ionizing fluxes expected to more accurately reproduce the emission line fluxes observed in star-forming galaxies.

Figure 2 shows the relative ionization fractions for a number of species produced by Mappings III, plotted as a function of relative distance from inner surface of the nebula. This figure illustrates that the [SII] flux in particular is a good tracer of the hardness of the ionizing radiation field. This is largely due to the effects of [SIII] ionization. In a soft ionizing radiation field [SIII] will be ionized out to a greater distance in the nebula, decreasing the ionization fraction of [SII] as a result. Conversely, in a hard ionizing radiation field, [SIII] will be ionized very close to the inner surface of the nebula and [SII] will dominate the ionization fraction. As a result, [SII] can be used as a powerful diagnostic tool in these models, particularly as it is a commonly-detected emission feature in star-forming galaxy spectra.

\section{Optical Emission Line Diagnostics}
\label{Sec-diag}
The final output of the Starburst99/Mappings III code is a model galaxy emission spectrum. The model spectra calculated for our grid of input parameters allows us to probe the ISM properties of these galaxies through the use of emission line diagnostics. In this work we employ the [NII]$\lambda$6584/H$\alpha$, [NII]$\lambda$6584/[OII]$\lambda$3727, [OIII]$\lambda$5007/H$\beta$, [OIII]$\lambda$5007/[OII]$\lambda$3727, and [SII]$\lambda\lambda$6717,6730/H$\alpha$ line ratios. As these fluxes can vary with the age of the young stellar population, we examine the evolution of each ratio as a function of time across our instantaneous burst SFH model grids, and consider this evolution in the context of the FUV ionizing spectra shown in Figure 1. For these ratios we assume an ionization parameter $q = 2 \times 10^8$ cm s$^{-1}$; this corresponds to log $\mathcal{U} \approx -2.2$, our closest ionization parameter to the log $\mathcal{U} = -2.3$ value adopted by Rigby \& Rieke (2004) in their models.

\subsection{[NII]/H$\alpha$}
The [NII] $\lambda6594$/H$\alpha$ ratio correlates strongly with both metallicity and ionization parameter, and is useful in diagnostic diagrams comparing these parameters (Veilleux \& Osterbrock 1987, Kewley et al.\ 2001). The [NII] flux at low metallicities is dominated by the abundance of primary nitrogen (Chiappini et al.\ 2005, Mallery et al.\ 2007). Primary nitrogen production is largely independent of metallicity and occurs predominantly in intermediate-mass stars (Matteucci \& Tosi 1985, Matteucci 1986); however, some primary nitrogen production by massive stars is thought to be dependent on stellar mass and metallicity (Chiappini et al.\ 2005, 2006; Mallery et al.\ 2007).  At higher metallicities the production of secondary nitrogen becomes prevalent (Alloin et al.\ 1979, Consid\`{e}re et al.\ 2000, Mallery et al.\ 2007); secondary nitrogen is synthesized from carbon and oxygen originally present in the star and is therefore proportional to abundance. At low metallicities, nitrogen abundance is also relatively sensitive to the star formation history of the galaxy, introducing some scatter into the relation between [NII]/H$\alpha$ and metallicity (Kewley \& Dopita 2002). The relatively low ionization potential of [NII], allowing production of this line at the lower-energy end of the FUV ionizing spectrum, also makes this ratio sensitive to ionization parameter (Kewley \& Dopita 2002).

The evolution of this ratio with age is shown in Figure 3a, where we plot all five metallicities and the HIGH and STD mass loss rates for the instantaneous burst SFH models. We can see an increase of the [NII]/H$\alpha$ ratio with metallicity. At the lowest metallicities ($Z = 0.05Z_{\odot}, Z = 0.2Z_{\odot}$) the [NII]/H$\alpha$ ratio increases by $\sim$80\%-160\% at later ages ($>$ 3 Myr), while at higher metallicities the ratio varies across all ages, showing deviations from the mean of up to $\sim$75\%-85\%. We can also see that [NII]/H$\alpha$ becomes double-valued at later ages ($>$ 4 Myr) for the highest metallicities ($Z = Z_{\odot}$ and $Z = 2Z_{\odot}$). To understand the variation of this ratio with age for an instantaneous burst SFH, we can consider the FUV ionizing spectrum. In the instantaneous burst FUV spectra, we see a larger variation with age at the ionizing threshold of [NII] as compared to the continuous burst FUV spectra, a consequence of the varying contribution from the coeval massive star population. This variation also becomes more prominent at higher metallicities, explaining the increased variation of the [NII]/H$\alpha$ ratio in that regime.

\subsection{[NII]/[OII]}
The [NII] $\lambda6594$ and [OII] $\lambda3727$ lines have very similar ionization potentials, making this ratio almost independent of the ionizing parameter of the radiation field. This ratio also avoids the pitfall of being double-valued with abundance, scaling smoothly from high to low abundance (van Zee et al.\ 1998) and rendering it a reliable means of isolating metallicity on diagnostic diagrams (Dopita et al.\ 2000). [NII]/[OII] correlates very strongly with metallicity above $Z = 0.4Z_{\odot}$; [NII]'s status as a secondary element causes it to more strongly scale with increasing metallicity than [OII]. In addition, the lower electron temperature at high metallicity produces fewer thermal electrons, decreasing the number of collisional excitations in the high-energy-threshold [OII] feature as compared to the lower-energy [NII] features (Kewley \& Dopita 2002). At lower metallicities ($Z \le 0.23Z_{\odot}$) this ratio's utility as a tracer of metallicity decreases slightly, since nitrogen and oxygen are {\it both} primary nucleosynthesis elements and begin to scale uniformly with metallicity as a result (Dopita et al.\ 2000).

The evolution of this ratio is shown in Figure 3b. This ratio shows increased variation with age at higher metallicities, similar to the behavior observed in the [NII]/H$\alpha$ ratio. As described in the [NII]/H$\alpha$ case, there is a variation with age in the FUV spectrum at the [NII] ionizing threshold that increases at higher metallicities, leading to increased variations in the flux of the [NII] line. In this case, we see the same behavior in the FUV spectrum at the ionizing threshold of [OII]. The slope of the FUV ionizing spectrum between these two thresholds increases with metallicity and age. This change in the relative ionizing flux between the two species leads to greater variation in the [NII]/[OII] flux at higher ages and higher metallicities. We can, however, see that the [NII]/[OII] ratio increases significantly with metallicity, demonstrating that on a diagnostic diagram a clear separation with metallicity will be evident across the [NII]/[OII] axis.

\subsection{[OIII]/H$\beta$}
The [OIII] $\lambda5007$/H$\beta$ ratio is sensitive to the hardness of the ionizing radiation field, and a useful means of tracing the ionizing parameter of a galaxy (Baldwin et al.\ 1981). However, this ratio is sensitive to metallicity and double-valued with respect to abundance, although it is far more sensitive to ionization parameter at sub-solar metallicities (Kewley et al.\ 2004). The close proximity of these emission lines also renders this ratio relatively insensitive to reddening corrections.

The evolution of the [OIII]/H$\beta$ ratio is shown in Figure 3c. We can see immediately that, as described in Kewley et al.\ (2004), [OIII]/H$\beta$ is double-valued with metallicity, reaching its highest value at $Z = 0.4Z_{\odot}$ and decreasing for both lower and higher metallicities. We can also see that the value of the ratio decreases with age. This decrease is gradual at the lowest metallicities but is interrupted by local maxima at 3 to 5 Myr as metallicity increases. This age range combined with the behavior of the FUV ionizing spectrum indicates that these local maxima in the [OIII]/H$\beta$ ratio are due to the short-lived contribution of the Wolf-Rayet phase to the flux of the [OIII] line. We can see at $Z = 0.2Z_{\odot}$ that this contribution from the Wolf-Rayet population is observed in the HIGH mass loss models but not the STD models; recall that at low metallicities in particular the HIGH mass loss models are expected to produce an earlier Wolf-Rayet population than the STD models. The overall decrease with age in the [OIII]/H$\beta$ flux can be attributed to the gradual decrease in hardness of the FUV ionizing spectrum at the ionization threshold of [OIII].

\subsection{[OIII]/[OII]}
The [OIII] $\lambda5007$/[OII] $\lambda3727$ serves a similar function as the [OIII]/H$\beta$ ratio, and is a commonly-employed ionization parameter diagnostic. Dopita et al.\ (2000) and Kewley \& Dopita (2002) do find that the [OIII]/[OII] is sensitive to metallicity as well, with higher abundances corresponding to lower values, but both the metallicity and ionization parameter relations are monotonic and do not result in this ratio being double-valued.

The evolution of the [OIII]/[OII] ratio is shown in Figure 3d, and supports the conclusions of Dopita et al.\ (2000); we can see that the value of this diagnostic ratio decreases as metallicity increases. The evolution of this diagnostic with age is quite similar to that of the [OIII]/H$\beta$ ratio. As in the case of [OIII]/H$\beta$, the [OIII] flux here is affected by the slowly decreasing hardness of the FUV spectra and the contributions of the Wolf-Rayet phase. By contrast, the flux of the FUV spectrum stays relatively constant at the ionizing threshold of [OII] for both SFHs, isolating the behavior of the [OIII] flux in the evolution of this ratio.

\subsection{[SII]/H$\alpha$}
The [SII] ($\lambda6716 + \lambda6731$)/H$\alpha$ line ratio is a useful means of tracing the hardness of the photoionizing spectrum present in a galaxy. While the ratio retains a slight dependence on the ionization parameter, we define this value at the innermost boundary of the nebula. [SII], by contrast, forms at a greater distance from this boundary in a partially ionized zone. As the hardness of the ionizing radiation field increases the partially ionized zone becomes more extended, generating increased [SII] flux.

The evolution of the [SII]/H$\alpha$ ratio is shown in Figure 3e. We can see that the value of this ratio increases gradually with abundance across the lower metallicities but becomes degenerate at high metallicities ($Z = Z_{\odot}, Z = 2Z_{\odot}$). We see that the [SII]/H$\alpha$ flux increases with age and undergoes large fluctuations at higher metallicities, showing a $\sim$90\%-95\% deviation from the mean. This is, once again, a signature of the increased variation with metallicity seen in the FUV ionizing spectrum at the [SII] ionizing threshold. Out of all of our diagnostic ratios, the [SII] flux is also the most sensitive to small variations in the hardness of the FUV ionizing spectrum.
\\
\\
\indent
We find that all the ratios are degenerate with age, particularly at high metallicity. It is therefore important to determine the 
young stellar population age for a galaxy before drawing any robust conclusions through the use of the instantaneous burst SFH models. We also find a measurable difference between the HIGH and STD mass loss rate models. This once again demonstrates that an instantaneous burst SFH traces the effect that a single coeval population of massive stars has on the FUV ionizing spectrum and clearly illustrates differences in the emission line spectrum imposed by the different stellar evolutionary tracks.

\section{Emission Line Diagnostic Diagrams}
We wish to test our Starburst99/Mappings models against observed spectra from a variety of galaxy populations. The local ($z < 0.1$) galaxies we selected for our comparison samples are described below.

{\it Sloan Digital Sky Survey (SDSS) galaxies}: To compare our models to the general local galaxy population, we plot emission line ratios determined for a sample of emission-line galaxies from SDSS described in Kewley et al.\ (2006). These galaxies were originally taken from Data Release 4 of SDSS (Adelman-McCarthy et al.\ 2006) and restricted to 85224 galaxies with a S/N $\ge$ 8 in the strong emission lines (Tremonti et al.\ 2004) and a redshift range between 0.04 $< z <$ 0.1. The lower limit of this redshift range corresponds to an aperture covering fraction of 20\%, the minimum required to avoid domination of the spectrum by aperture effects (Kewley et al.\ 2005, Kewley \& Ellison 2008). The spectra were acquired with 3" diameter fibers. We further restrict this sample to the 60920 galaxies classified as star-forming by Kewley et al.\ (2006). This classification is based on the criteria of Kauffmann et al.\ (2003) derived from the [NII]/H$\alpha$ vs. [OIII]/H$\beta$ diagnostic diagram, and the Kewley et al.\ (2001) criteria based on the [SII]/H$\alpha$ vs. [OIII]/H$\beta$ and [OI]/H$\alpha$ vs. [OIII]/H$\beta$ diagnostic diagrams; equations are given in Kewley et al.\ (2006). This classification removes contaminations from composite galaxies, Seyferts, LINERs, and ambiguous galaxies that cannot be definitively classified based on the Kauffmann et al.\ (2003) and Kewley et al.\ (2001) criteria.

{\it Nearby Field Galaxy Survey (NFGS) galaxies}: To ensure that the local galaxy population comparison is not affected by any residual aperture effects present in the SDSS sample, we include a sample of galaxies from NFGS. NFGS has integrated emission-line spectrophotometry of 196 galaxies that span from $-14 < M_B < -22$ and include the full range of Hubble sequence morphologies, with a median redshift of $z = 0.01$ and a maximum redshift of $z \sim 0.07$ (Jansen et al.\ 2000a, 2000b). The galaxies were selected from the CfA redshift catalog, which has a limiting blue photographic magnitude of $m_Z = 14.5$ (Huchra et al.\ 1983) and observed with the FAST spectrograph at the Whipple 1.5 m telescope (Jansen et al.\ 2000b). We limit this sample to 95 star-forming galaxies by applying the emission-line criteria for star-forming galaxies in Kewley et al.\ (2006) described above for the SDSS sample.

{\it Blue Compact Galaxies (BCGs)}: To compare our models to a sample of galaxies that show evidence of burst-like star formation histories, we include a sample of blue compact galaxies (BCGs) from Kong \& Cheng (2002) and Kong et al.\ (2002). Their sample consists of 97 galaxies selected from the of Gordon \& Gottesman (1981), Thuan \& Martin (1981), Kinney et al.\ (1993), and Thuan et al.\ (1999) and limited to $M_B < -17$ mag. The spectra were observed with the OMR spectrograph on the 2.16m telescope of the Beijing Astronomical Observatory. Here we restrict our sample to those galaxies with a complete set of spectral line fluxes that satisfy the star-forming galaxy emission line criteria in Kewley et al.\ (2006), giving us 36 star-forming BCGs in our final sample. Note that, for consistency with our other star-forming galaxy samples, we use the Kewley et al.\ (2006) criteria rather than the classifications used in Kong et al.\ (2002) to remove AGN and non-emission-line galaxies from their sample. This sample of BCGs has a redshift range of $0.003 < z < 0.029$, with a median redshift of $z = 0.016$.

{\it Metal-Poor Galaxies (MPGs)}: Finally, to examine the agreement of our stellar population synthesis and photoionization models with low metallicity galaxies, we include a sample of 10 MPG spectra from Brown et al.\ (2008). The 10 MPG spectra are part of a survey of 38 MPG candidates, selected from Data Release 4 of SDSS based on restrictions on their ($u' - g'$)$_0$, ($g' - r'$)$_0$, and ($r' - i'$)$_0$ colors, a magnitude limit of $g' < 20.5$, and removal of HII regions in nearby galaxies through visual inspection. The remaining MPG candidates were observed with the Blue Channel spectrograph at the MMT telescope. Brown et al.\ (2008) find 10 MPGs in this sample, with log(O/H) + 12 $<$ 8 based on their electron temperature metallicities (Izotov et al.\ 2006). These 10 galaxies have a redshift range of $ 0.03 < z < 0.081$ with a median redshift of $z = 0.073$.

All of the emission line fluxes used in our comparison samples have been corrected for total intrinsic attenuation by dust in the direction of the galaxy. This was based on the H$\alpha$/H$\beta$ emission line ratio, assuming the Balmer decrement for case B recombination (H$\alpha$/H$\beta$ = 2.87 for T = 10$^4$ K and $n_e$ $\sim$ 10$^2$ - 10$^4$ cm$^{-3}$, following Osterbrock 1989) and the Cardelli et al.\ (1989) reddening law with the standard total-to-selective extinction ratio $R_V$ = 3.1.

These comparisons samples allow us to examine the agreement of our stellar population synthesis and photoionizaion models when applied to the local galaxy population, including nearby low-metallicity galaxies. We plot these comparison samples along with our models on a series of optical emission line ratio diagnostic diagrams, utilizing the ratios detailed in Section 3.

\subsection{[NII]/H$\alpha$ vs. [OIII]/H$\beta$}
In Section 3, we note that [NII]/H$\alpha$ correlates strongly with metallicity as well as with ionization parameter (Veilleux \& Osterbrock 1987, Kewley et al.\ 2001; see Section 3.1), while [OIII]/H$\beta$ is primarily a measure of ionization parameter with a degenerate dependence on metallicity (Baldwin et al.\ 1981, Kewley et al.\ 2004; see Section 3.3). The [NII]/H$\alpha$ vs. [OIII]/H$\beta$ grid of Baldwin et al.\ (1981) allows us to examine these two ISM properties by examining the evolution of the grids with age and comparing the agreement of our models to observed galaxy spectra. In Figure 4 we plot the evolution of the [NII]/H$\alpha$ vs. [OIII]/H$\beta$ diagnostic grid for our instantaneous burst SFH models ranging from 0.0 to 5.0 Myr, while Figure 5 shows the diagnostic diagram assuming a continuous SFH and an age of 5.0 Myr; both figures compare the model grid to the SDSS, NFGS, BCG, and MPG samples described above.

Following the observed evolution of these diagnostic ratios in Section 3, we can see that the instantaneous burst models in Figure 4 show a substantial decrease in the [OIII]/H$\beta$ flux with age, attributable to the significant decrease in the hardness of the FUV spectrum with age. When comparing these grids to our galaxy sample, we see satisfactory agreement for the youngest instantaneous burst models (0.0 Myr, 1.0 Myr); these models accommodate 69\% of the SDSS galaxies, 78\% of the NFGS sample, 58\% of the BCGs, and 86\% of the MPGs, a great improvement over previous model grids that show poor agreement with metal-poor galaxies. However, this agreement rapidly deteriorates with age as the [OIII]/H$\beta$ ratio in the models decreases and the fluctuating contributions of the Wolf-Rayet phase are introduced at 3.0 to 5.0 Myr. The MPG sample in particular is in poor agreement with the models at 3.0 to 5.0 Myr. At later ages (4.0 Myr, 5.0 Myr) the HIGH mass loss models show slightly better agreement with the data as compared to the STD models. By comparison, we find that the continuous SFH models in Figure 5 show a better agreement with the data at 5.0 Myr, although deficiencies in the model [OIII]/H$\beta$ fluxes are apparent.

From this diagnostic, it {\it appears} that our sample population is restricted to the higher metallicities included in our models (with the MPGs and a small sample of outliers as the only exception) and spans the full range of ionization parameters included in our models, again with only a small sample of outliers. However, the double-valued nature of this diagnostic makes it an impractical sole means of drawing strong conclusions about galaxies' metallicities and ionization parameters. 

\subsection{[NII]/[OII] vs. [OIII]/[OII]}
In Figure 6 we plot the evolution of [NII]/[OII] vs. [OIII]/[OII] with age for the instantaneous burst SFH models and compare them to our galaxy samples; in Figure 7 we again plot the 5.0 Myr continuous SFH models. We can see from the model grids plotted in Figure 6 and Figure 7 that [NII/OII] is primarily sensitive to metallicity, while [OIII/OII] is primarily sensitive to ionization parameter, following from discussion of these diagnostic ratio in Sections 3.2 and 3.4.

This diagnostic shows much more consistent agreement with our samples - the best-fit model grids at 0.0Myr accommodate 91\% of the SDSS galaxies, 66\% of the NFGS sample, 86\% of the BCGs, and 86\% of the MPGs. For the instantaneous burst models, we see a rapid decrease with age in the [OIII]/[OII] ratio and fluctuations from the Wolf-Rayet phase contribute at 3.0 to 5.0 Myr. The 1.0 to 2.0 Myr instantaneous burst models also appear to be somewhat deficient at higher metallicities in the case of the SDSS sample, while the 5.0 Myr grid shows poor agreement with the low-metallicity MPG sample. At 5.0 Myr, the HIGH mass loss models are also found to be in better agreement with our galaxy samples than the STD models. By contrast, the continuous SFH models show consistent agreement (66\%-95\%) with our galaxy samples at 5.0 Myr for both the HIGH and STD mass loss evolutionary track models, including all but the most metal-poor galaxy in the MPG sample.

In this non-degenerate diagnostic we can see that the SDSS galaxies span the full range of metallicities adopted in our models, although they appear to be restricted to the lower ionization parameters ($q \le 8 \times 10^7$ cm s$^{-1}$). The NFGS and BCG samples span a similar range in ionization parameter, but do not extend to the twice-solar metallicity regime of our models. Interestingly, the MPG sample appear to coincide with both the lowest metallicities ($Z = 0.05Z_{\odot}$ to $Z = 0.2Z_{\odot}$) {\it and} the highest ionization parameters present across our comparison samples, along with several of the most metal-poor BCGs, ranging from $q = 8 \times 10^7$ cm s$^{-1}$ to $q = 2 \times 10^8$ cm s$^{-1}$.

\subsection{[SII]/H$\alpha$ vs. [OIII]/H$\beta$}
In Figure 8 we plot the evolution of the [SII]/H$\alpha$ vs. [OIII]/H$\beta$ diagnostic ratio for our models ranging from 0.0 to 5.0 Myr assuming an instantaneous burst SFH, while Figure 9 shows the diagnostic at 5.0 Myr for a continuous SFH. In both figures we compare the model grid to our galaxy samples.

Both the instantaneous burst and continuous SFH models show partial agreement with the galaxy samples; however, it is clear in these diagrams that the grid is not able to accommodate a large number of these galaxies. For the best-fit instantaneous burst models at 0.0 Myr, the models only show agreement with 61\% of the SDSS galaxies, 26\% of the NFGS galaxies, 33\% of the BCGs, and 50\% of the MPGs. Agreement is better across the [SII]/H$\alpha$ axis, but due to the double-valued nature of the diagram we cannot isolate the behavior of this line ratio. In the instantaneous burst models, we see fluctuations in the [OIII]/H$\beta$ ratio with age and progressively poorer agreement with the galaxy samples, beginning with a failure to accommodate any of the galaxies in the MPG sample by 3.0 Myr but progressing to only agreeing with 35\% of the SDSS galaxies, 13\% of the BCGs, and 6\% of the NFGS galaxies by 5.0 Myr for the HIGH mass loss grid; the STD mass loss grid shows a $\sim$0\% agreement with our galaxy samples at 5.0 Myr. The continuous SFH models maintain a somewhat more satisfactory agreement at 5.0 Myr and show little distinction between the HIGH and STD mass loss rates, but the grids are still insufficient and show on 30\% agreement with the MPG sample. As discussed in Section 3.5, use of the [SII]/H$\alpha$ diagnostic allows us to examine the hardness of the FUV ionizing spectrum; Figures 8 and 9 suggest that a generally harder FUV ionizing spectrum is required from the models to produce a more extended partially ionized zone in the theoretical nebula.

\subsection{Comparison with Previous Model Grids}
The stellar population synthesis and photoionization models presented in Dopita et al.\ (2006) are comparable in many ways to the models presented in this work. They use the Starburst99 stellar population synthesis code, adopting the Pauldrach et al.\ (2001) and Hillier \& Miller (1998) model atmospheres that include treatments of metal opacities and the five metallicities of the Geneva HIGH evolutionary tracks. They also use the latest generation of Mappings III to compute their photoionization models. The Dopita et al.\ (2006) models adopt a spherical geometry in the photoionization models, while this work assumes a plane-parallel nebular geometry - these different Mappings geometries have been compared and produce equivalent results (Kewley et al.\ 2001). There are, however, two noteworthy differences between the models of Dopita et al.\ (2006) and those presented here.

First, the models of Dopita et al.\ (2006) do not take ionization parameter ($q$) as one of their free parameters; instead, $q$ is replace by $\mathcal{R}$, a parameter representing the ratio of the mass of the central aging star cluster to the pressure of the surrounding ISM. This allows $q$ to vary with age (as well as with metallicity). In this treatment, $\mathcal{R}$ is fixed at a variety of values ($-6,-4, -2, 0,$ and 2) and a sequence of model HII regions are computed at each of the Geneva metallicities with ages that are increased in increments of 0.5Myr, up to a maximum age of 6.5Myr. 

Second, Dopita et al.\ (2006) generate models of individual HII regions. To model the spectra of star-forming galaxies, they integrate the fluxes of a model HII region for each of their model ages, essentially considering star-forming galaxies to have spectra that consist of contributions from multiple HII regions at different ages. This removes age as a free parameter in these models, restricting the free parameters to metallicity and $\mathcal{R}$ (similar to our models with free parameters of metallicity and $q$). 

As compared to the Kewley et al.\ (2001) models, the primary difference in this work lies in the choice of model atmospheres. While this work adopts the the Pauldrach et al.\ (2001) and Hillier \& Miller (1998) model atmospheres, the Kewley et al.\ (2001) grids utilize the Lejeune et al.\ (1997) and Schmutz et al.\ (1992) models, which do not include treatments of metal opacities. Kewley et al.\ (2001) cite this as a shortcoming of their models and suggest that  the inclusion of continuum metal opacities will result in a harder FUV spectrum, a prediction that is supported by our models (see Section 3). The Kewley et al.\ (2001) models also include slightly different treatments of $n_e$ ($n_e$ = 350 cm$^{-3}$ as compared to the $n_e$ = 100 cm$^{-3}$ models shown here) and age (adopting a continuous SFH model age of 8.0 Myr as opposed to the 5.0 Myr used in this work). However, both of these differences are expected to have negligible effects on the diagnostic grids produced by these models.

In Figure 10 (left) we compare our 0.0 Myr instantaneous burst SFH models to the Kewley et al.\ (2001) and Dopita et al.\ (2006) models on the emission line diagnostic diagrams described above, considering the agreement of all three grids with the SDSS and MPG galaxy samples. The Kewley et al.\ (2001) models also assume a 0.0 Myr instantaneous burst SFH and range from $q = 1 \times 10^7$ to $q \ 3 \times 10^8$; the Dopita et al.\ (2006) models assume an instantaneous burst SFH and range from 0.0 Myr to 4.0 Myr ($q$ is not a free parameter in the Dopita et al.\ 2006 models). For the [NII]/H$\alpha$ vs. [OIII]/H$\beta$ diagnostic diagram (top left), our models are found to more closely agree with the data than either the Kewley et al.\ (2001) or the Dopita et al.\ (2006) grids. The Dopita et al.\ (2006) do not produce sufficient [OIII] fluxes at higher metallicities. The Kewley et al.\ (2001) accommodate all of the galaxies but do not properly track the empirical Kaufmann et al.\ (2003) cut-off for star-forming galaxies that has been applied to the SDSS sample, suggesting that the [OIII] fluxes produced by these models are unrealistically high. By contrast, the model grids produced by this work track the Kaufmann et al.\ (2003) cut-off perfectly. In the case of the [NII]/[OII] vs. [OIII]/[OII] (center left), all three models show agreement with the SDSS and MPG samples, although the agreement with the MPG sample is better for the Dopita et al.\ (2006) models and this work (6 out of 7) than for the Kewley et al.\ (2001) models (3 out of 7). Finally, the [SII]/H$\alpha$ vs. [OIII]/H$\beta$ diagram (bottom left) shows the failure of the Dopita et al.\ (2006) models to produce accurate [SII]/H$\alpha$ ratios for either sample. The Kewley et al.\ (2001) models have higher [OIII] fluxes than this work; however, our models extend to higher values of [SII]/H$\alpha$ in the SDSS sample. Across all three of these models, we find that changes in the treatment of dust have minimal effects in the optical regime; however, this is expected to make a more substantial difference in the infrared.

This comparison is noteworthy when considering that the models presented here are found to be an improvement over, or at least equivalent to, the integrated model spectra of Dopita et al.\ (2006). In our work we are modeling galaxy environments in a simple way by assuming the HII regions observed within an aperture can be modeled by a luminosity-weighted mean HII region represented by a plane parallel model. It is significant that this simple approach produces comparable results when compared with the more sophisticated treatment of modeling the integrated spectra of multiple HII regions employed in the Dopita et al.\ (2006) models. We can therefore conclude that treating model galaxies as single luminosity-weighted HII regions, particularly when assume a zero-age instantaneous burst SFH, is a simpler and equally effective approach in stellar population synthesis and photoionization modeling.

In Figure 10 (right) we compare our 5.0 Myr continuous SFH models to the Kewley et al.\ (2001) 8.0 Myr continuous SFH models, which again range from $q = 1 \times 10^7$ to $q \ 3 \times 10^8$. For the [NII]/H$\alpha$ vs. [OIII]/H$\beta$ diagnostic diagram (top right), we can see that the agreement of our models with the Kaufmann et al.\ (2003) cut-off in the SDSS sample has degraded for these later-age models; the Kewley et al.\ (2001) maintain higher [OIII] fluxes. In the [NII]/[OII] vs. [OIII]/[OII] diagnostic diagram (center right), we again see a slightly improved agreement with the MPG sample for our models (6 out of 7) as compared to the Kewley et al.\ (2001) models (4 out of 7); our models appear to extend to lower metallicities for the [NII]/[OII] diagnostic ratio. Finally, on the [SII]/H$\alpha$ vs. [OIII]/H$\beta$ diagnostic diagram (bottom right) we can see that both model grids do a poor job of accommodating the SDSS and MPG samples, although the Kewley et al.\ (2001) once again has higher [OIII] fluxes as compared to the models in this work.

This result is surprising considering the speculation by Kewley et al.\ (2001) that the inclusion of model atmospheres with metal opacities should lead to a harder FUV spectrum and strong line fluxes. While we do indeed produce a harder FUV spectrum by including the Pauldrach et al.\ (2001) and Hillier \& Miller (1998) models in our Starburst99 simulations, the effect on the line ratios produced by Mappings III appears to be less than anticipated. Our models do include a more accurate treatment of physical conditions, and show improvements in the agreement with the Kaufmann et al.\ (2003) criteria for star-forming galaxies and the metal-poor galaxies along the metallicity-sensitive [NII]/[OII] diagnostic ratio. However, it is clear that more substantial improvements in models of star-forming galaxies will require further systematic changes in existing stellar population synthesis and photoionization codes. 

\section{Discussion and Future Work}
\label{Sec-disc}
We have generated an extensive suite of models for star-forming galaxies, utilizing the newest generation of the Starburst99 stellar population synthesis code and the Mappings III photoionization code. With these codes we have constructed a grid of models with a variety of metallicities and ionization parameters, adopting both an instantaneous burst SFH and a continuous SFH as well as both the HIGH and STD mass loss prescriptions from the evolutionary tracks of the Geneva group. This grids have been made available to the public\footnotemark \footnotetext{The models are available for download at \texttt{http://www.emlevesque.com/model-grids/}}, and are being used to develop updated metallicity diagnostics as part of a current HST theory grant (P.I. Kewley), with a publication of new diagnostics expected within the next few months.

We have examined the ionizing spectrum generated by Starburst99 for these models, along with the evolution of a number of optical emission line diagnostic ratios with time. We have selected a number of local ($z < 0.1$) star-forming galaxies as a comparison sample, from Data Release 4 of SDSS (Kewley et al.\ 2006), the NFGS survey of Jansen et al.\ (2000a, 2000b), the blue compact galaxy survey of Kong \& Cheng (2002) and the MPG galaxy sample of Brown et al.\ (2008). By comparing our models to these data on a series of emission line diagnostic diagrams, we are able to make the following conclusions:

(1) Models that assume a continuous SFH at 5.0 Myr produce a harder FUV ionizing spectrum and show better agreement with the observed emission line ratios of our galaxy sample as compared to models with an instantaneous burst SFH at 5.0 Myr. This is in accordance with past work that suggests a continuous treatment of star formation is more realistic than a single zero-age instantaneous burst when modeling star-forming galaxies (Kewley et al.\ 2001, Moy et al.\ 2001, Ferndandes et al.\ 2003, Noeske et al.\ 2007a, 2007b).

(2) Assumption of either the HIGH or STD Geneva mass loss rates is found to make very little difference in the precision of the continuous SFH models; however, in the case of the instantaneous burst models, the HIGH mass loss rates produce better agreement with our galaxy sample, suggesting that an enhanced rate of mass loss is more realistic under the assumption of an instantaneous burst.

(3) From the [NII]/H$\alpha$ vs. [OIII]/H$\beta$ and [NII]/[OII] vs. [OIII]/[OII] diagnostic diagrams, we find that the metallicity and ionization parameter range of our models are in agreement with the ISM properties of the galaxies in our comparison sample, including our low-metallicity galaxy sample. We find that that these metal-poor galaxies appear to include both lower metallicities {\it and} higher ionization parameters than our general local galaxy sample.

(4) From the [SII]/H$\alpha$ vs. [OIII]/H$\beta$ diagnostic diagram, it appears that our models still produce an insufficiently hard FUV ionizing spectrum that cannot fully reproduce some of the line fluxes observed in our galaxy sample.

(5) Our models of a single luminosity-weighted HII region are comparable in precision to models such as those presented in Dopita et al.\ (2006), which integrate the spectra of multiple HII regions when modeling star-forming galaxies.

It is important to note that these models still include several shortcomings that must be considered when applying them to star-forming galaxies and considering future work in this area. One ongoing challenge we must consider is the required hardness of the FUV ionizing spectrum. The Starburst99 stellar population synthesis models shown here have a considerably harder FUV ionizing spectrum in the 225\AA\ to 1000\AA\ regime than the Kewley et al.\ (2001) models, a regime that is critical in ionizing the forbidden optical emission lines used in these analyses. This is a consequence of adopting the Pauldrach et al.\ (2001) and Hillier \& Miller (1998) model atmospheres, which include detailed treatments of metal opacities, an improvement originally proposed in Kewley et al.\ (2001). However, our results show that the models still do not produce sufficient flux in the FUV ionizing spectrum; this is most evident in the case of the [SII]/H$\alpha$ vs. [OIII]/H$\beta$ diagnostic. [SII] in particular requires a larger partially ionized zone generated by a harder radiation field in the models before it can properly reproduce the fluxes observed in our galaxy samples. This suggests that further systematic changes are required in the stellar population synthesis models to produce harder FUV ionizing spectra.

One potential means of addressing this issue could be the adoption of stellar evolutionary tracks that include the effects of rotation on the stellar population, such as the new generation of Geneva evolutionary tracks presented in V\'{a}zquez et al.\ (2007). Starburst99 outputs generated using the $z = 0.02$ rotating Geneva models have been made available to us (Leitherer, personal communication), allowing us to examine the effect that rotation has on the ionizing radiation field produced by this code. Figure 11 shows a comparison of the FUV ionizing spectra generated by Starburst99 when adopting the HIGH, STD, and rotating models of the Geneva group; it is evident that evolutionary tracks that include rotation produce an ionizing spectrum that is in general harder than either of the tracks employed in this work, with the difference becoming most pronounced at wavelengths shorter than $\sim$225\AA. This is precisely the effect that would be anticipated as a result of including rotation in massive stellar evolutionary models. As just one example of the changes expected with these new tracks, massive stars are found to be hotter and more luminous than previously thought, leading to a hardening of the SED that is most prominent in the higher-energy regime of the spectrum (Leitherer et al.\ 2008). In particular, these rotation effects are expected to become more pronounced at lower metallicity, as rotation eventually becomes a dominant parameter in the evolution of extremely metal-poor stars (Hirschi et al.\ 2008, Leitherer et al.\ 2008). We plan to expand our current grid of evolutionary models by generating a new set of Starburst99/Mappings III synthetic spectra once the full grid of Geneva evolutionary tracks with rotation are available.

In summary, we find that our stellar population synthesis and photoionization models of a single luminosity-weighted HII region, adopting a continuous SFH, reproduce the line ratios of up to 91\% of the local population of star-formation galaxies, as well as up to 86\% of the low-metallicity population. These models will allow us to study the ISM properties of metal-poor galaxies in unprecedented detail, allowing study of the pre-enrichment ISM that could shed new light on star formation processes and mechanisms of ISM enrichment (Brown et al.\ 2008). Models of low-metallicity galaxies can also be used to probe potential evolutionary mechanisms for metal-poor environments (Kewley et al.\ 2007. Finally, a thorough understanding of metal-poor galaxies and their stellar populations could prove beneficial to studying the host galaxies of long-duration gamma-ray bursts, which are thought to be low-metallicity (Stanek et al.\ 2006, Fruchter et al.\ 2006, Kewley et al.\ 2007, Modjaz et al.\ 2008). These models still include several shortcomings that must be considered, in particular their production of an insufficiently hard FUV ionizing spectrum that specifically affects the [SII] emission line strength. Future model grids that include the effects of rotation on the stellar population may help to resolve this issue.

We would like to thank the anonymous referee for extremely helpful and constructive comments regarding this manuscript. We gratefully acknowledge our useful correspondence with Warren Brown, Margaret Geller, Claus Leitherer, Georges Meynet, Daniel Schaerer, and Leonie Snijders. E. Levesque's participation was made possible in part by a Ford Foundation Predoctoral Fellowship. K. Larson's participation in this project was made possible in part through the National Science Foundation's Research Experience for Undergraduates program. L. Kewley and E. Levesque gratefully acknowledge support by NSF EARLY CAREER AWARD AST07-48559.

\begin{figure}[h]
\includegraphics[width=8cm]{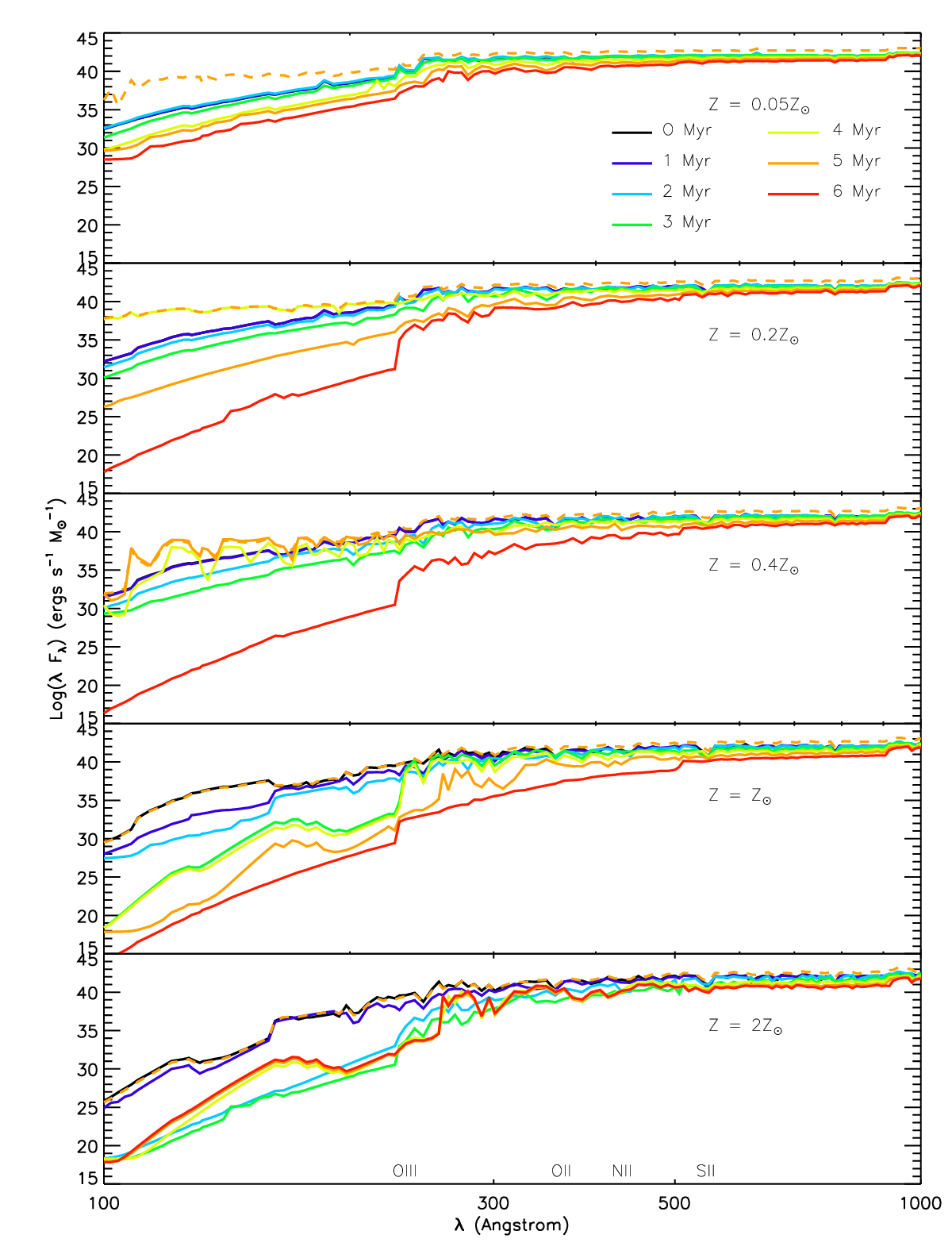}
\includegraphics[width=8cm]{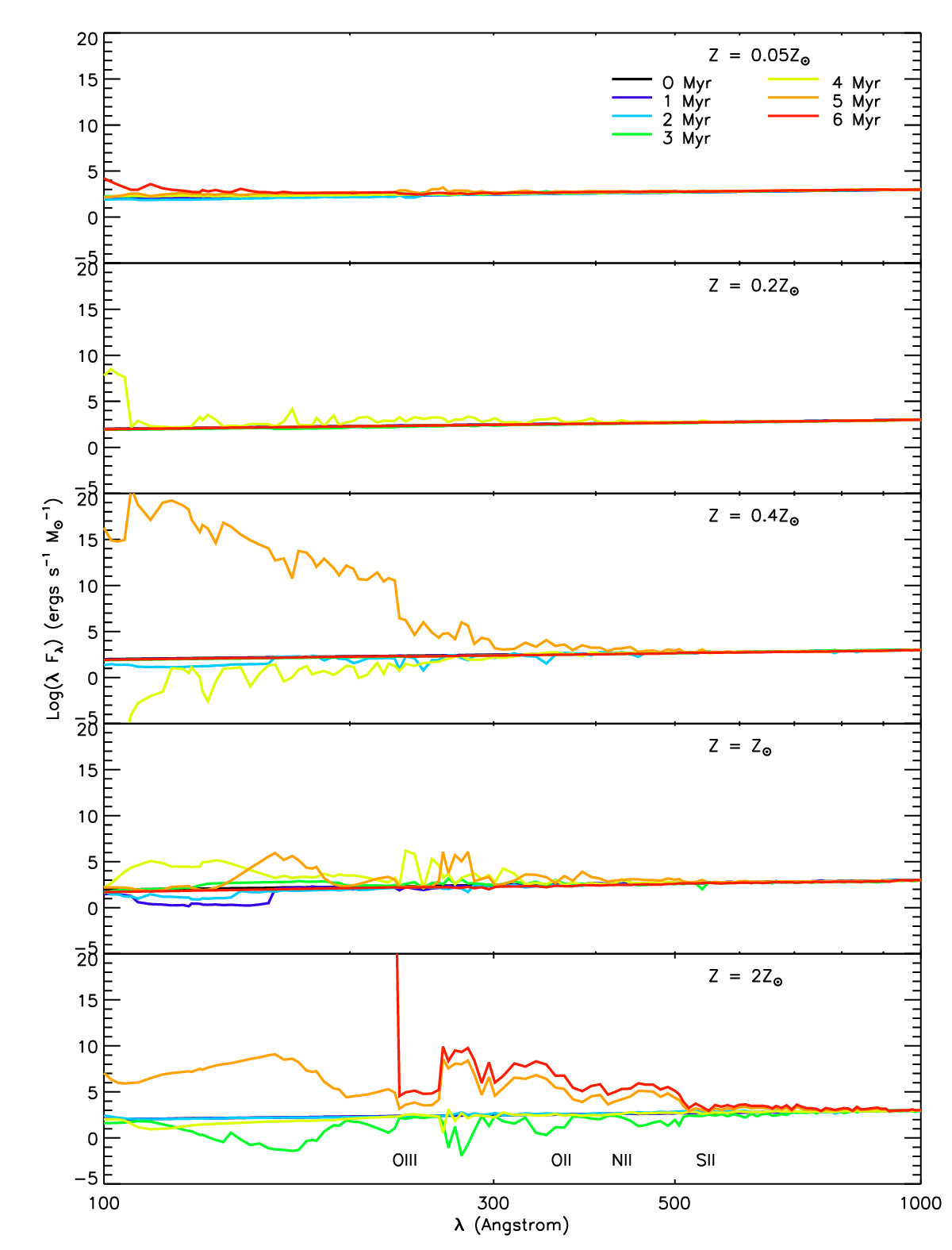}
\caption{{\it Left:} FUV spectra generated by the Starburst99 code adopting an instantaneous burst SFH. The spectra were generated using the Geneva HIGH evolutionary models, showing the progression of the spectra with age in 1 Myr increments for the full range of metallicities. At 5 Myr the FUV spectra generated by Starburst99 assuming an continuous SFH are also shown (dotted line). Ionization potentials for the relevant elements are marked on the x-axis. {\it Right:} Subtraction of the FUV spectra when adopting the Geneva HIGH evolutionary tracks and the STD evolutionary tracks; HIGH $-$ STD is plotted. In both panels the wavelengths are plotted on a log scale.}
\end{figure}
\clearpage

\begin{figure}[h]
\includegraphics[width=17cm]{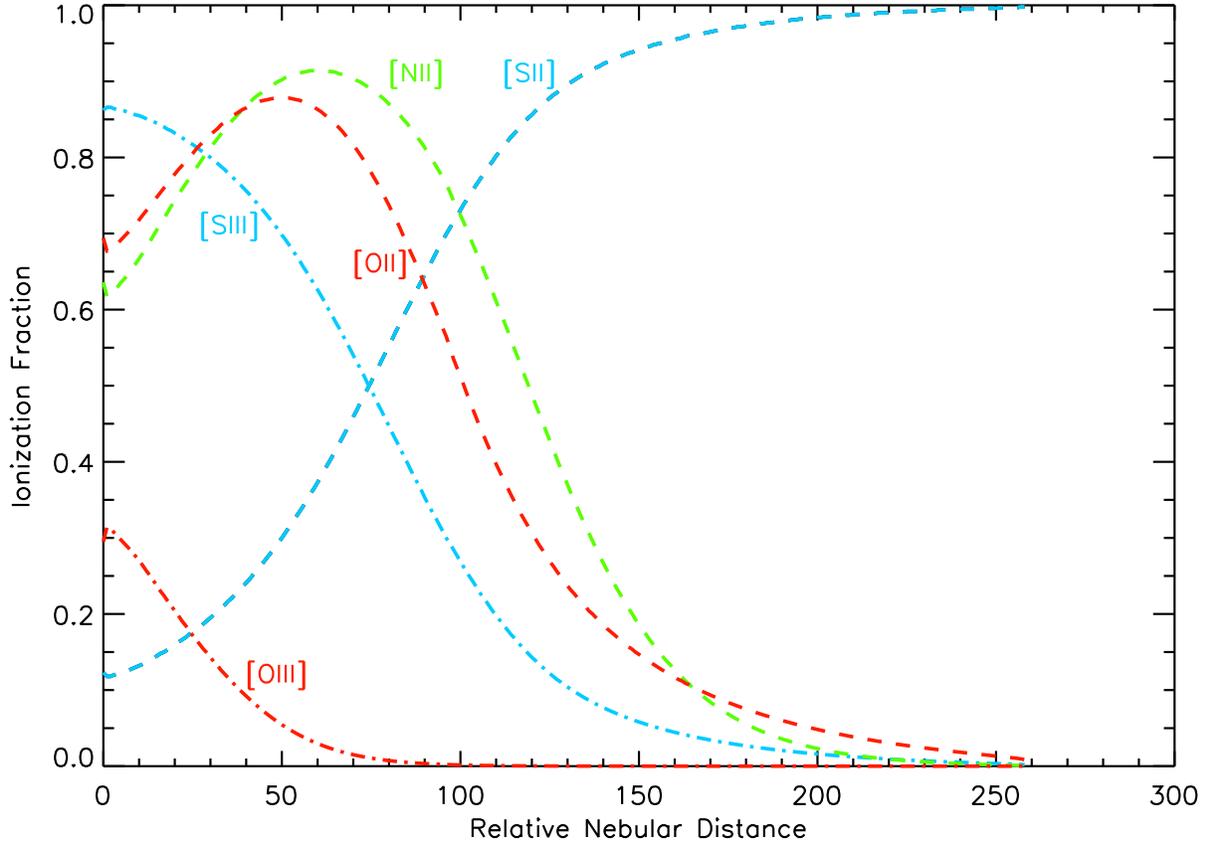}
\caption{Relative ionization fractions for emission lines produced by Mappings III, plotted as a function of relative distance from inner surface of the nebula. The emission features shown here are [NII] (dashed green line), [OII] (dashed red line), [OIII] (dashed-dotted red line), [SII] (dashed blue line), and [SIII] (dashed-dotted blue line). For these outputs a zero-age instantaneous burst SFH, a metallicity of Z = 0.05$Z_{\odot}$, an ionization parameter $q = 1 \times 10^7$, and an electron density $n_e = 100$ cm$^{-3}$ is assumed.}
\end{figure}

\begin{center}
\begin{figure}[h]
\label{ratios}
\includegraphics[width=9cm]{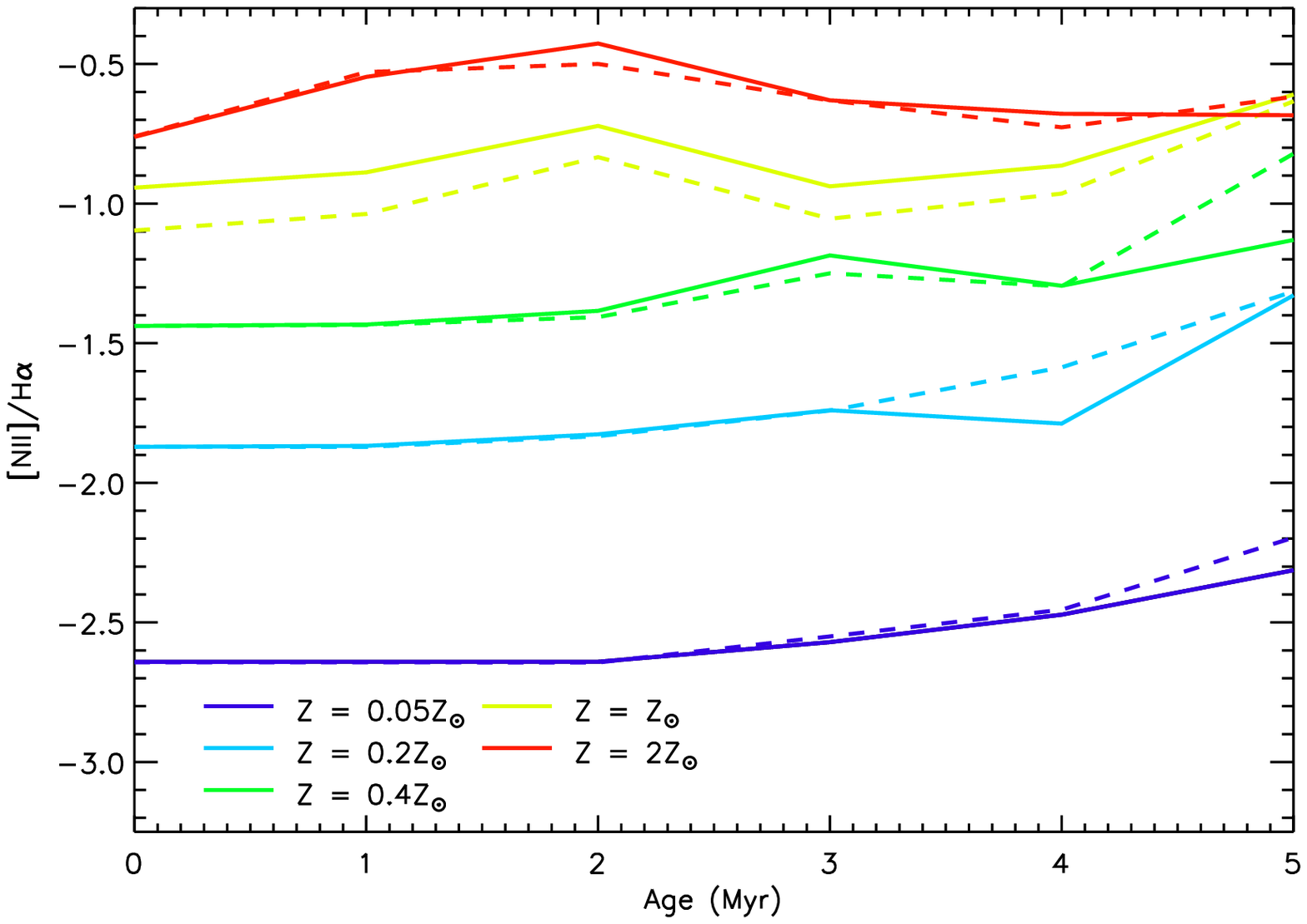}
\includegraphics[width=9cm]{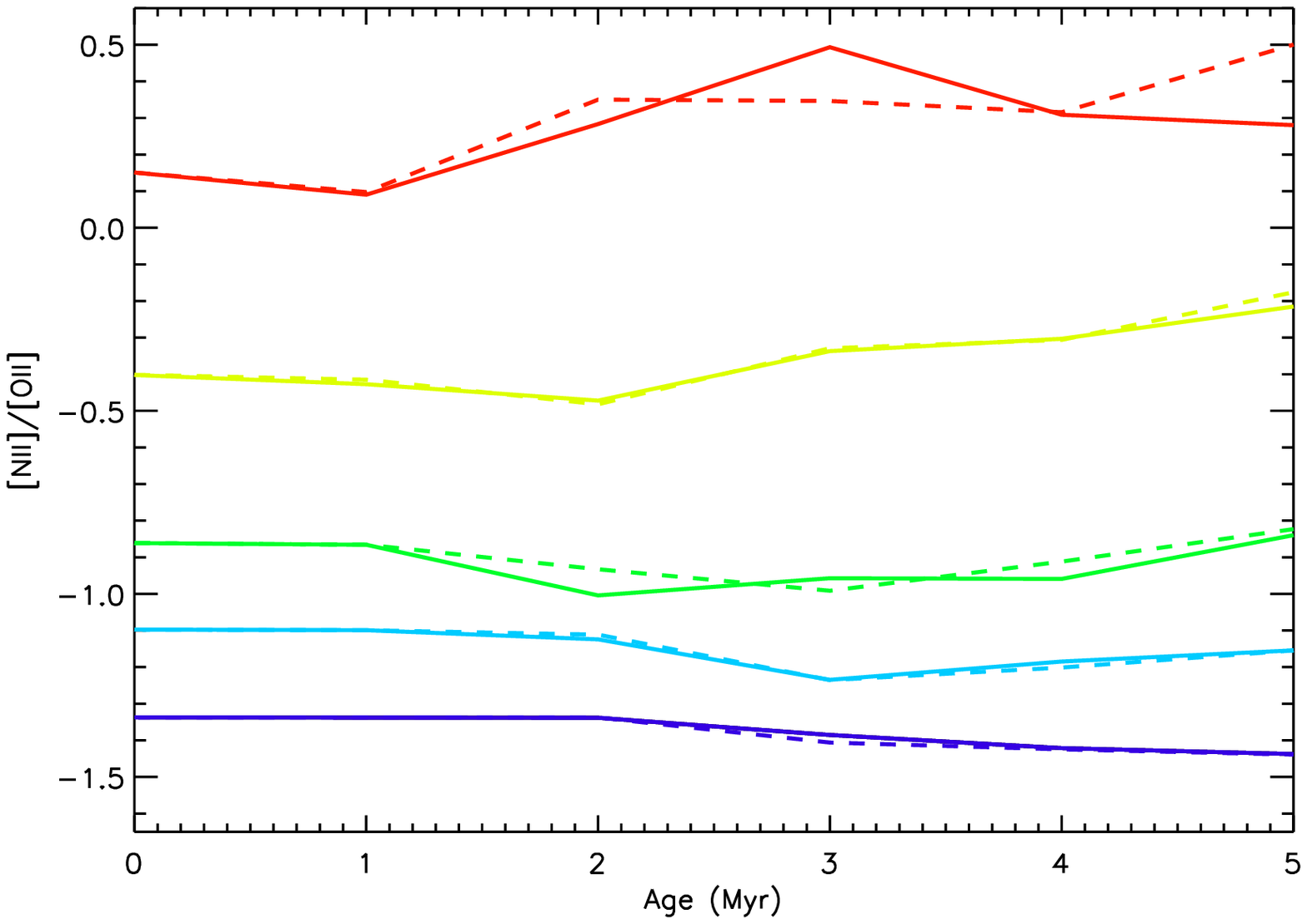}
\includegraphics[width=9cm]{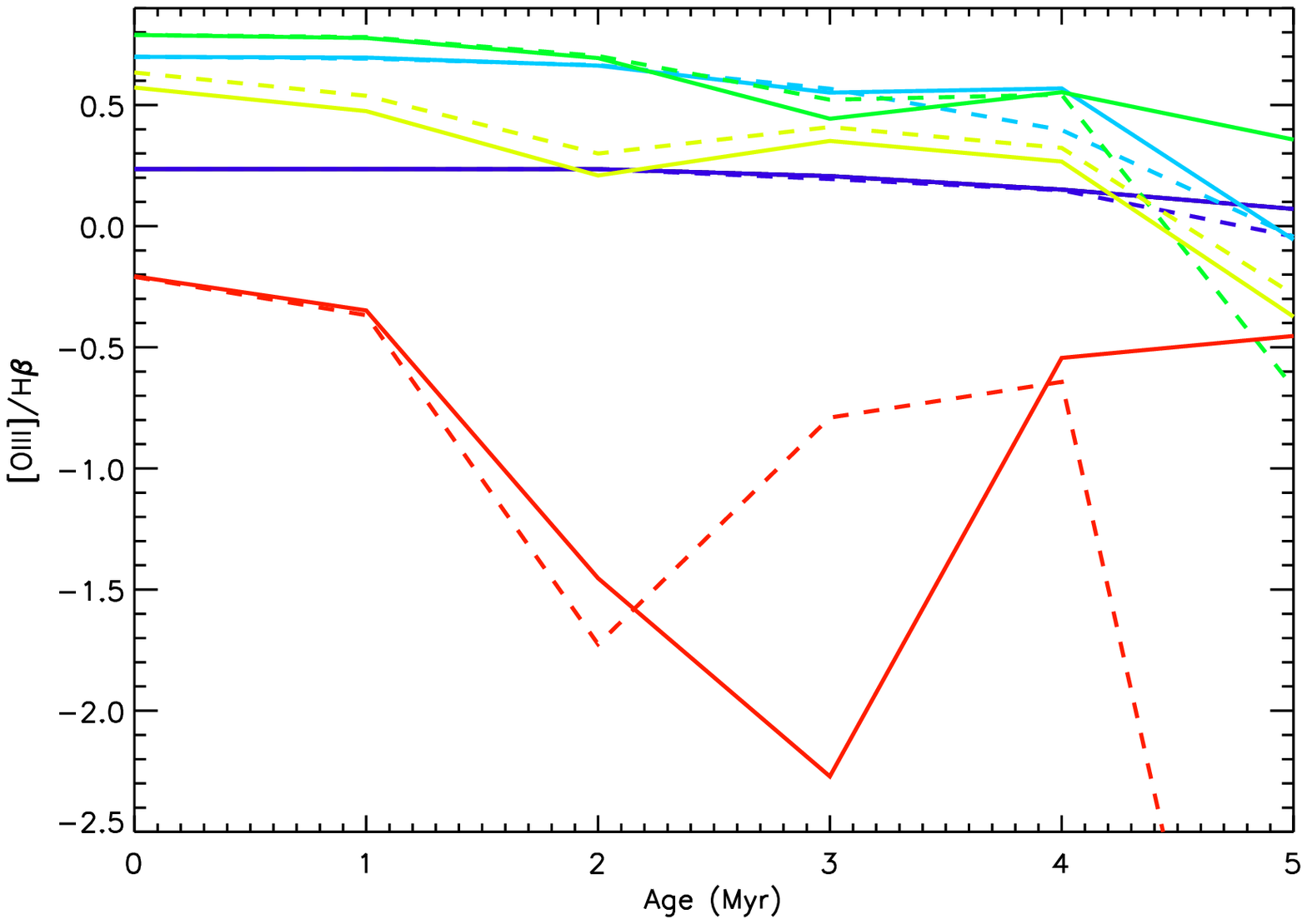}
\includegraphics[width=9cm]{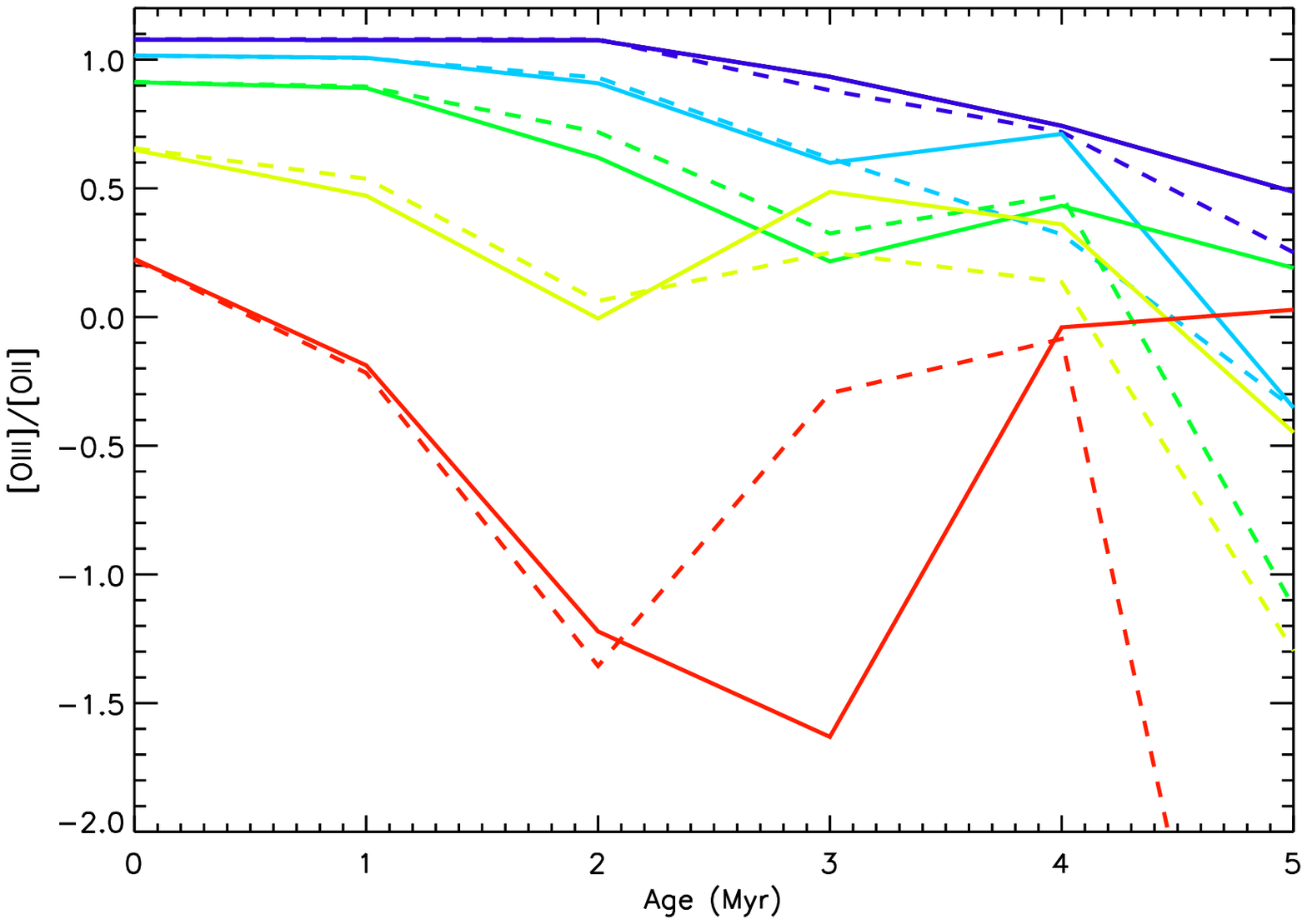}
\includegraphics[width=9cm]{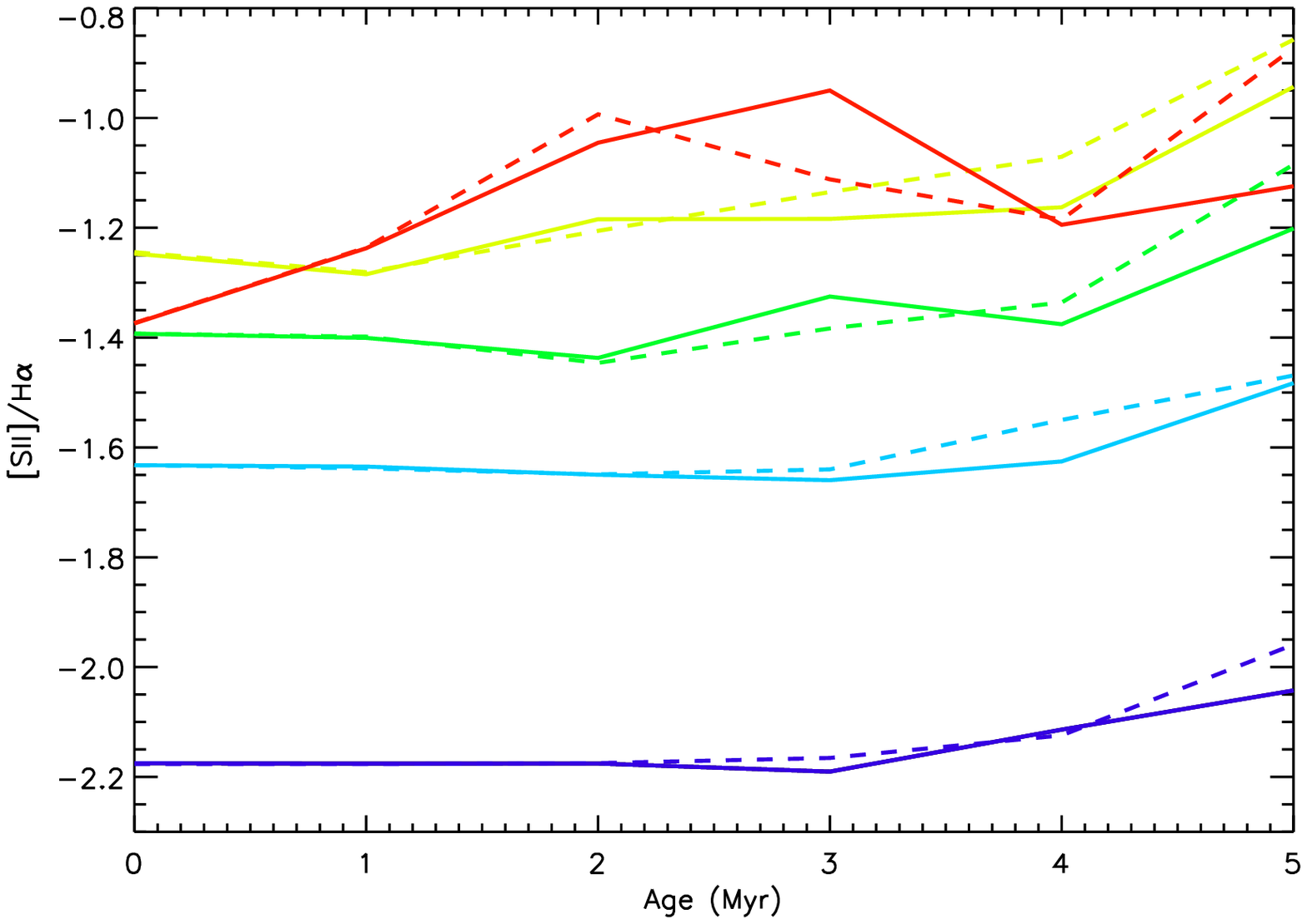}
\caption{Evolution of the diagnostic emission line ratios with age, shown for all five metallicies and ranging from 0 to 5 Myr assuming an instantaneous burst star formation history. Models generated with the Geneva HIGH (solid line) and Geneva STD (dashed line) evolutionary tracks are compared. An $n_e$ = 100 is assumed.}
\end{figure}
\end{center}

\begin{figure}[h]
\begin{center}
\includegraphics[width=18cm]{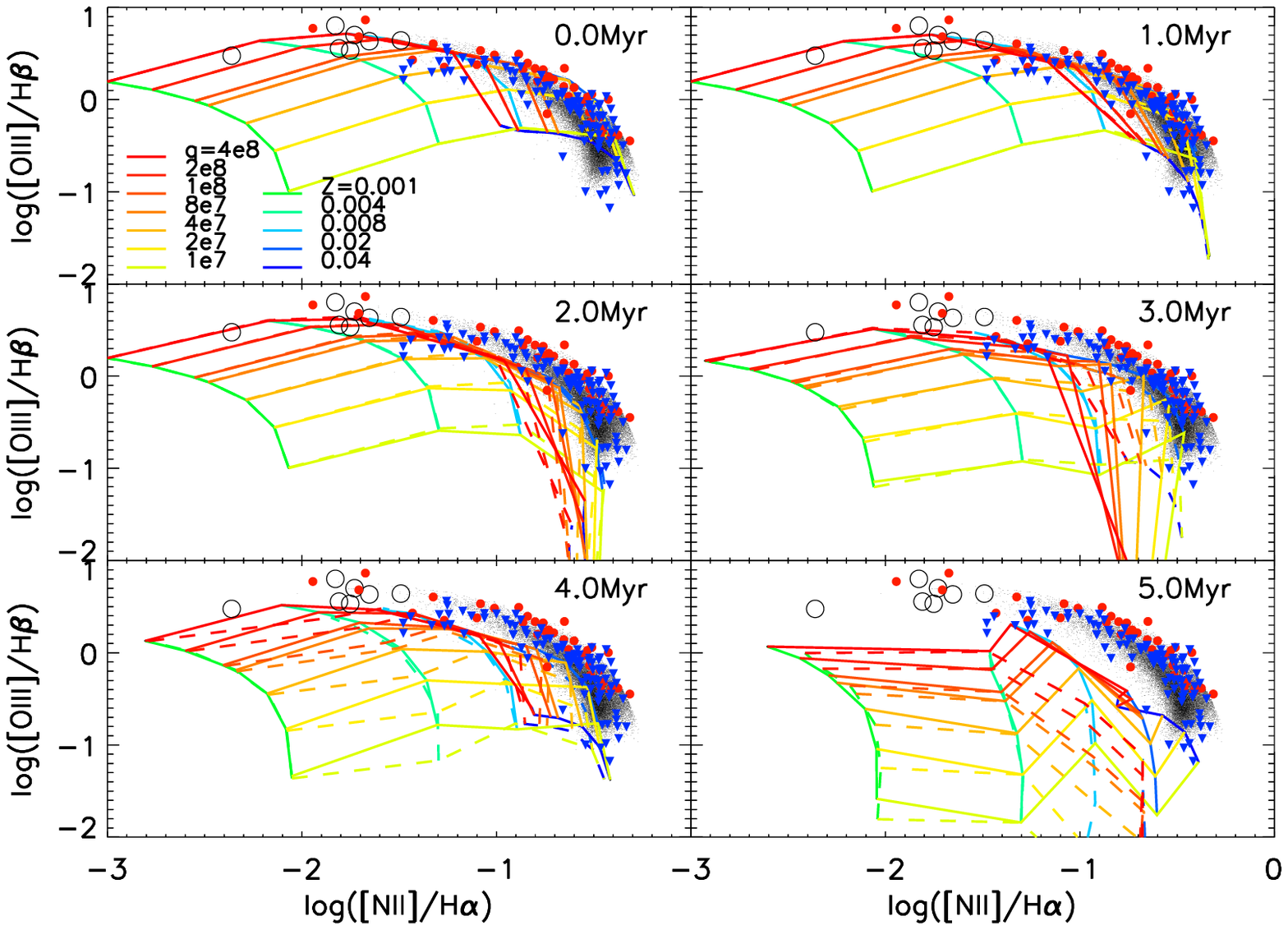}
\caption{[NII]/H$\alpha$ vs. [OIII]/H$\beta$ diagnostics for the instantaneous burst SFH model grids evolving from 0.0 Myr to 5.0 Myr in increments of 1.0 Myr. The models are plotted with lines of constant metallicity vs. lines of constant ionization parameter. Grids generated with the Geneva HIGH evolutionary tracks are plotted with solid lines, while grids generated with the Geneva STD tracks are plotted with dashed lines. An electron density $n_e = 100$ is assumed. The grids are compared to our sample of 60,920 SDSS star-forming galaxies from Kewley et al.\ (2006) (points), 95 NFGS galaxies from Jansen et al.\ (2006b) (blue triangles), blue compact galaxies from Kong \& Cheng (2002) (red circles), and 10 metal-poor galaxies from Brown et al.\ (2006) (large open circles).}
\end{center}
\end{figure}

\begin{figure}[h]
\begin{center}
\includegraphics[width=18cm]{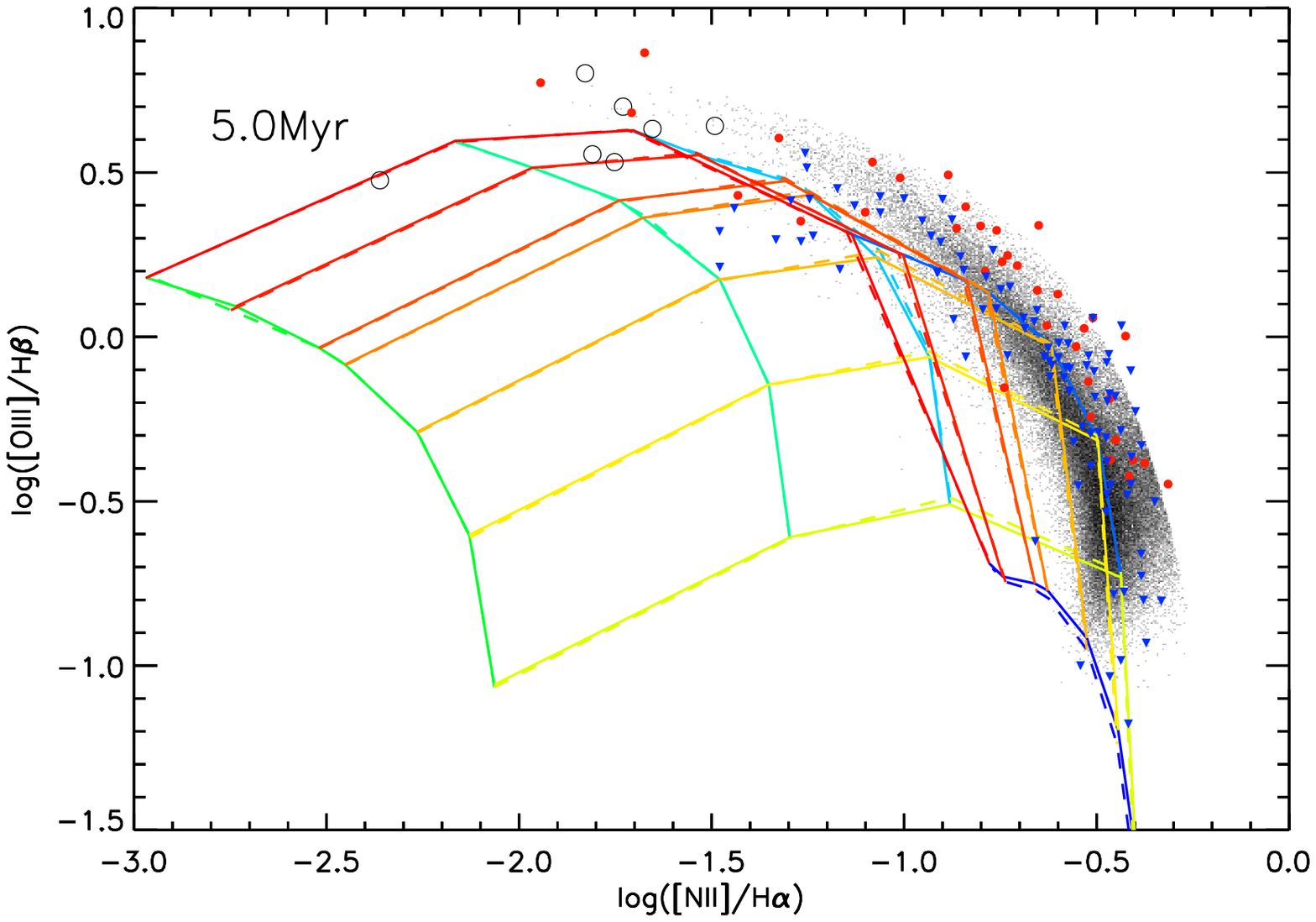}
\caption{[NII]/H$\alpha$ vs. [OIII]/H$\beta$ diagnostics for the continuous SFH model grids at an age of 5.0 Myr. The models are plotted with lines of constant metallicity vs. lines of constant ionization parameter. Grids generated with the Geneva HIGH evolutionary tracks are plotted with solid lines, while grids generated with the Geneva STD tracks are plotted with dashed lines. An electron density $n_e = 100$ is assumed. The grids are compared to our sample of 60,920 SDSS star-forming galaxies from Kewley et al.\ (2006) (points), 95 NFGS galaxies from Jansen et al.\ (2006b) (blue triangles), blue compact galaxies from Kong \& Cheng (2002) (red circles), and 10 metal-poor galaxies from Brown et al.\ (2006) (large open circles). Lines of constant metallicity and ionization parameter follow the legend shown in Figure 4.}
\end{center}
\end{figure}

\begin{figure}[h]
\begin{center}
\includegraphics[width=18cm]{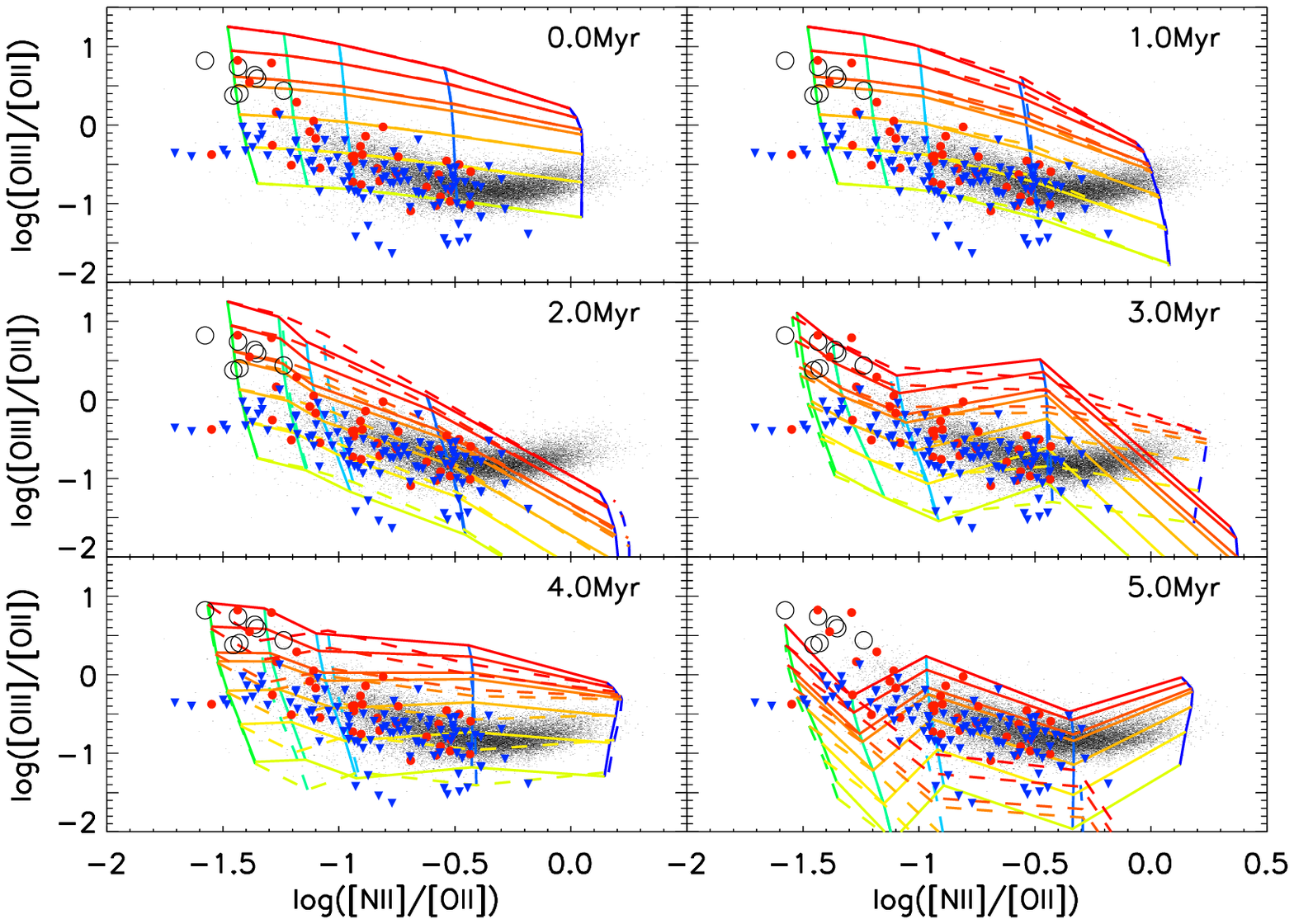}
\caption{[NII]/[OII] vs. [OIII]/[OII] diagnostics for the instantaneous burst SFH model grids evolving from 0.0 Myr to 5.0 Myr in increments of 1.0 Myr. The models are plotted with lines of constant metallicity vs. lines of constant ionization parameter. Grids generated with the Geneva HIGH evolutionary tracks are plotted with solid lines, while grids generated with the Geneva STD tracks are plotted with dashed lines. An electron density $n_e = 100$ is assumed. The grids are compared to our sample of 60,920 SDSS star-forming galaxies from Kewley et al.\ (2006) (points), 95 NFGS galaxies from Jansen et al.\ (2006b) (blue triangles), blue compact galaxies from Kong \& Cheng (2002) (red circles), and 10 metal-poor galaxies from Brown et al.\ (2006) (large open circles). Lines of constant metallicity and ionization parameter follow the legend shown in Figure 4.}
\end{center}
\end{figure}

\begin{figure}[h]
\begin{center}
\includegraphics[width=18cm]{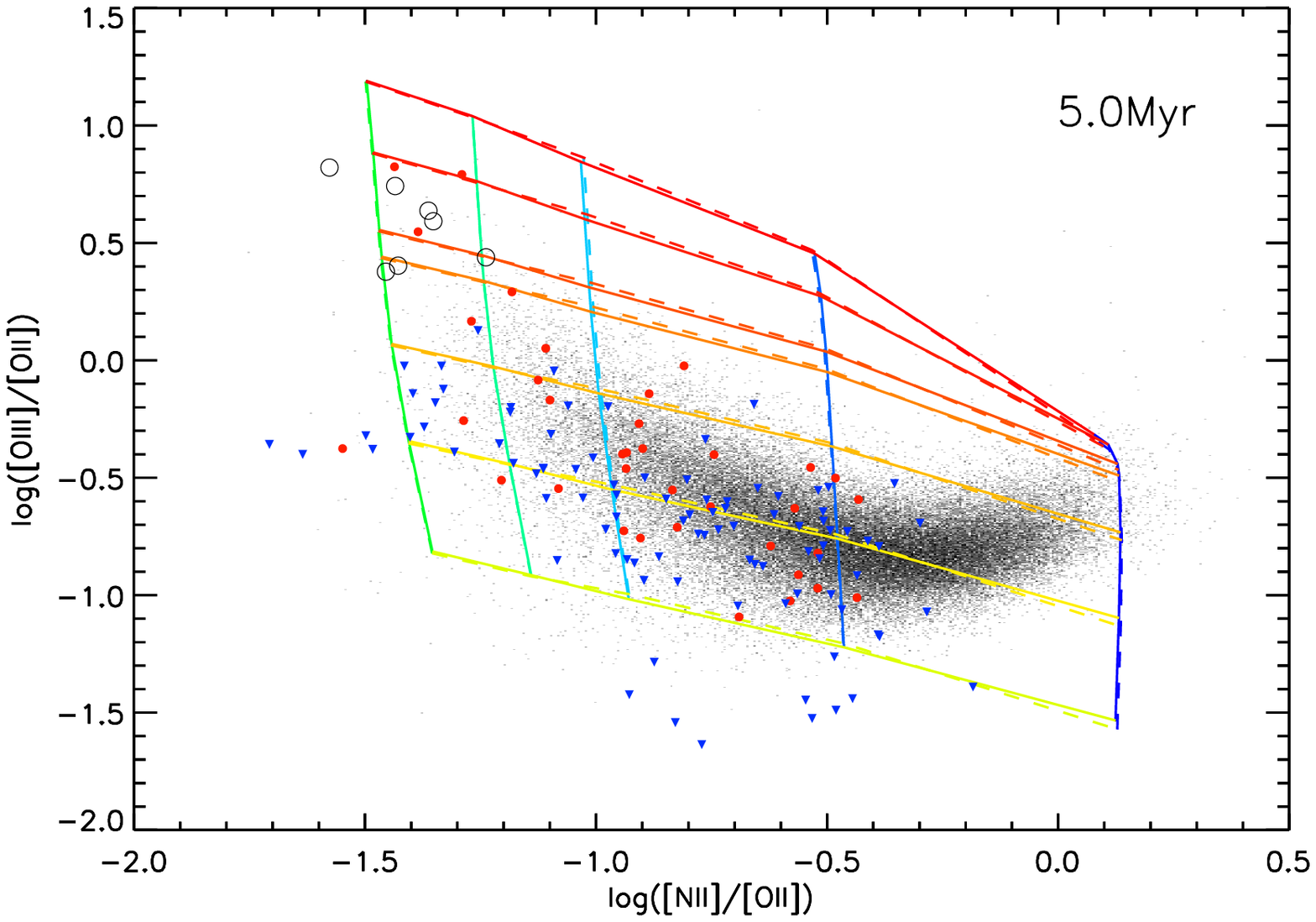}
\caption{[NII]/[OII] vs. [OIII]/[OII] diagnostics for the continuous SFH model grids at an age of 5.0 Myr. The models are plotted with lines of constant metallicity vs. lines of constant ionization parameter. Grids generated with the Geneva HIGH evolutionary tracks are plotted with solid lines, while grids generated with the Geneva STD tracks are plotted with dashed lines. An electron density $n_e = 100$ is assumed. The grids are compared to our sample of 60,920 SDSS star-forming galaxies from Kewley et al.\ (2006) (points), 95 NFGS galaxies from Jansen et al.\ (2006b) (blue triangles), blue compact galaxies from Kong \& Cheng (2002) (red circles), and 10 metal-poor galaxies from Brown et al.\ (2006) (large open circles). Lines of constant metallicity and ionization parameter follow the legend shown in Figure 4.}
\end{center}
\end{figure} 

\begin{figure}[h]
\begin{center}
\includegraphics[width=18cm]{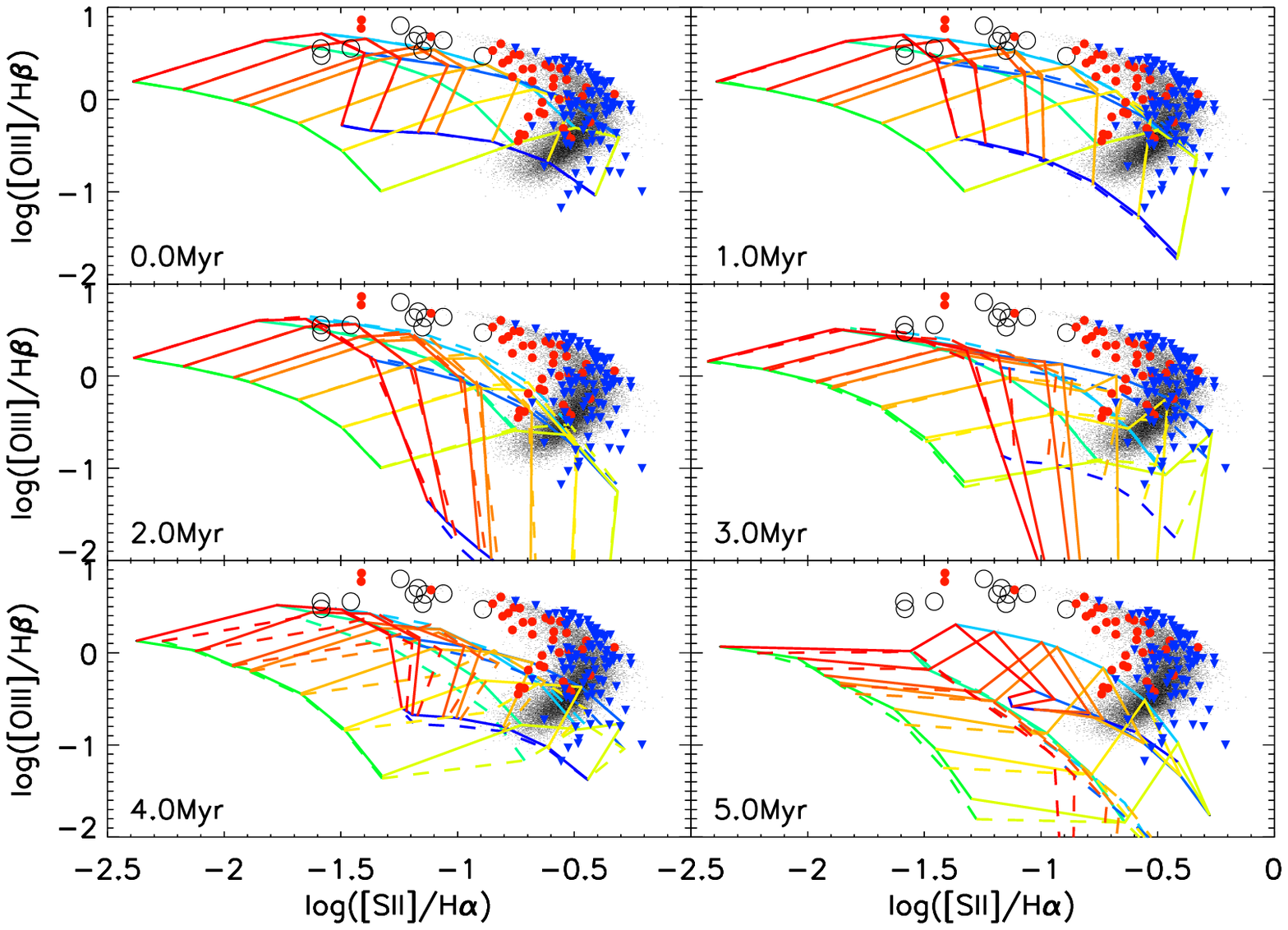}
\caption{[SII]/H$\alpha$ vs. [OIII]/H$\beta$ diagnostics for the instantaneous burst SFH model grids evolving from 0.0 Myr to 5.0 Myr in increments of 1.0 Myr. The models are plotted with lines of constant metallicity vs. lines of constant ionization parameter. Grids generated with the Geneva HIGH evolutionary tracks are plotted with solid lines, while grids generated with the Geneva STD tracks are plotted with dashed lines. An electron density $n_e = 100$ is assumed. The grids are compared to our sample of 60,920 SDSS star-forming galaxies from Kewley et al.\ (2006) (points), 95 NFGS galaxies from Jansen et al.\ (2006b) (blue triangles), blue compact galaxies from Kong \& Cheng (2002) (red circles), and 10 metal-poor galaxies from Brown et al.\ (2006) (large open circles). Lines of constant metallicity and ionization parameter follow the legend shown in Figure 4.}
\end{center}
\end{figure}

\begin{figure}[h]
\begin{center}
\includegraphics[width=18cm]{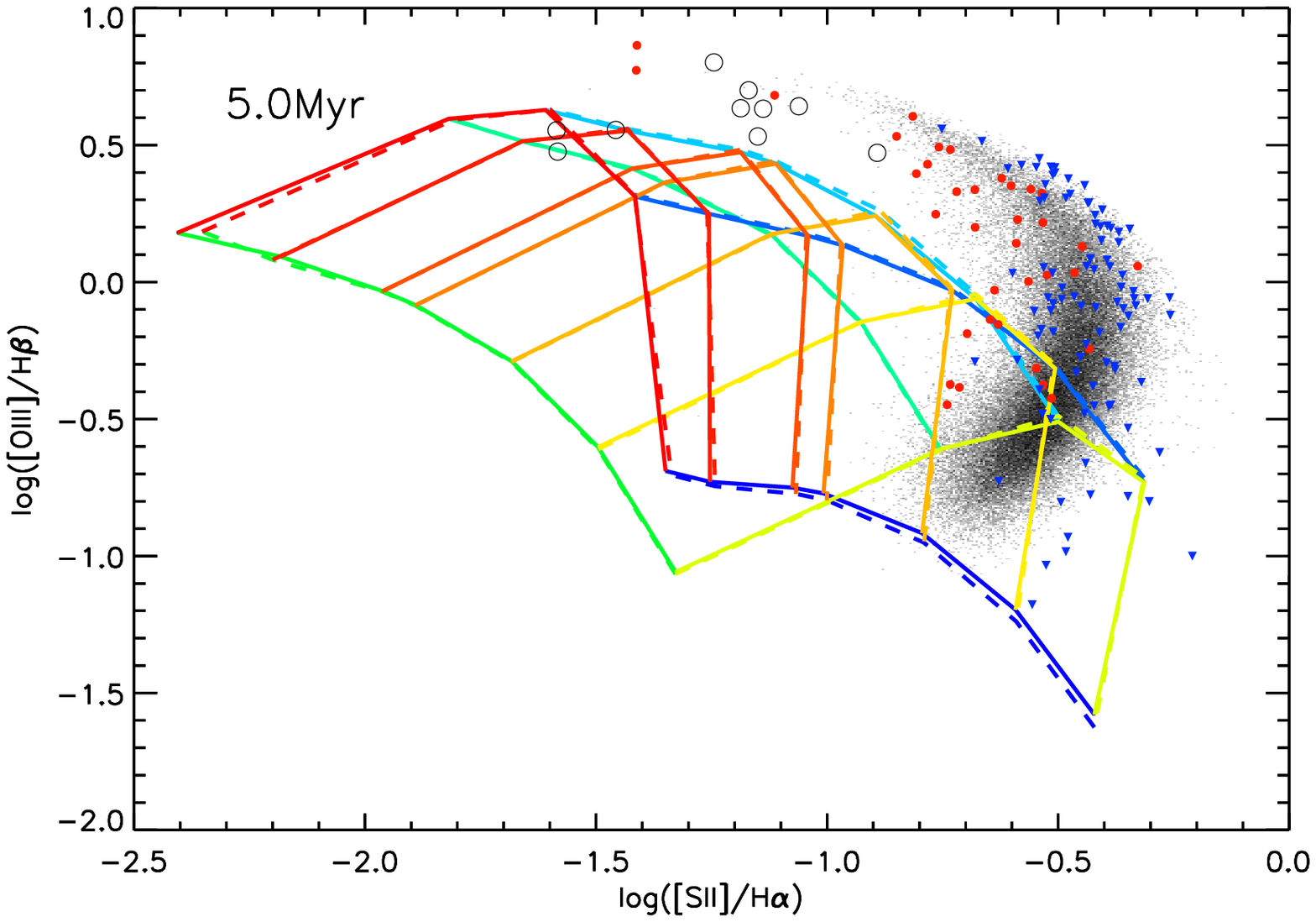}
\caption{[SII]/H$\alpha$ vs. [OIII]/H$\beta$ diagnostics for the continuous SFH model grids at an age of 5.0 Myr. The models are plotted with lines of constant metallicity vs. lines of constant ionization parameter. Grids generated with the Geneva HIGH evolutionary tracks are plotted with solid lines, while grids generated with the Geneva STD tracks are plotted with dashed lines. An electron density $n_e = 100$ is assumed. The grids are compared to our sample of 60,920 SDSS star-forming galaxies from Kewley et al.\ (2006) (points), 95 NFGS galaxies from Jansen et al.\ (2006b) (blue triangles), blue compact galaxies from Kong \& Cheng (2002) (red circles), and 10 metal-poor galaxies from Brown et al.\ (2006) (large open circles). Lines of constant metallicity and ionization parameter follow the legend shown in Figure 4.}
\end{center}
\end{figure}

\begin{figure}
\centering
\includegraphics[width=8.5cm]{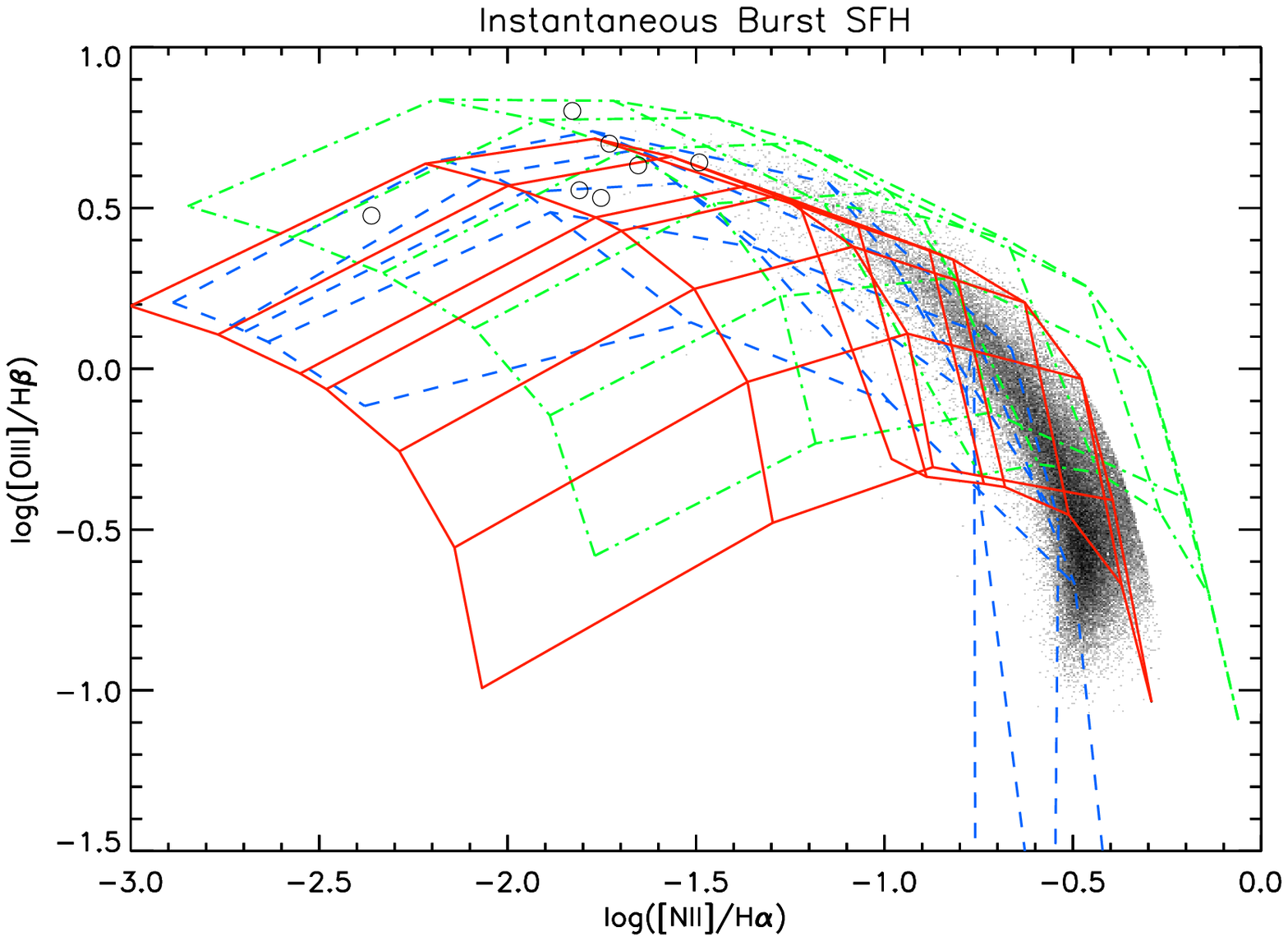}
\includegraphics[width=8.5cm]{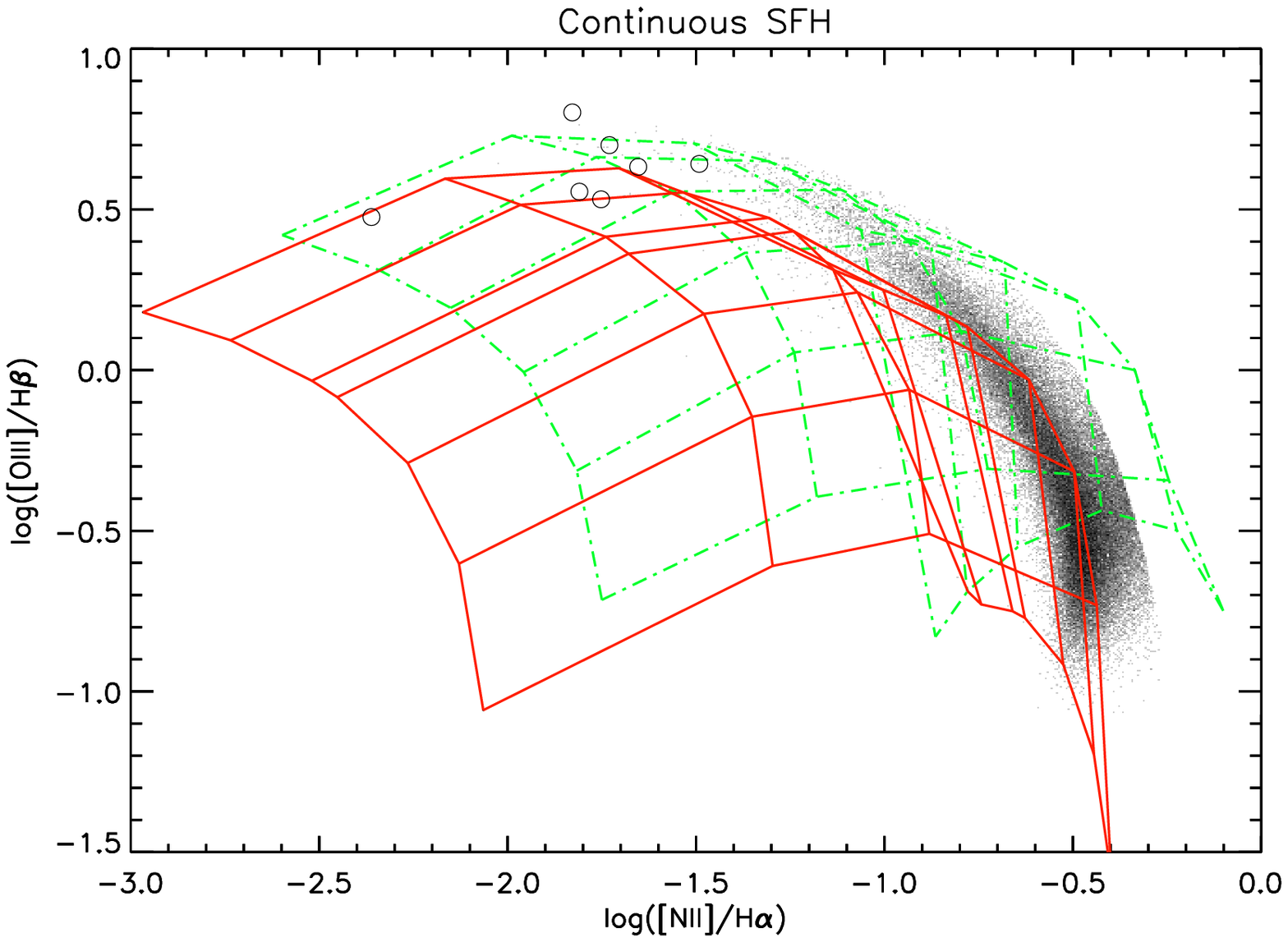}
\includegraphics[width=8.5cm]{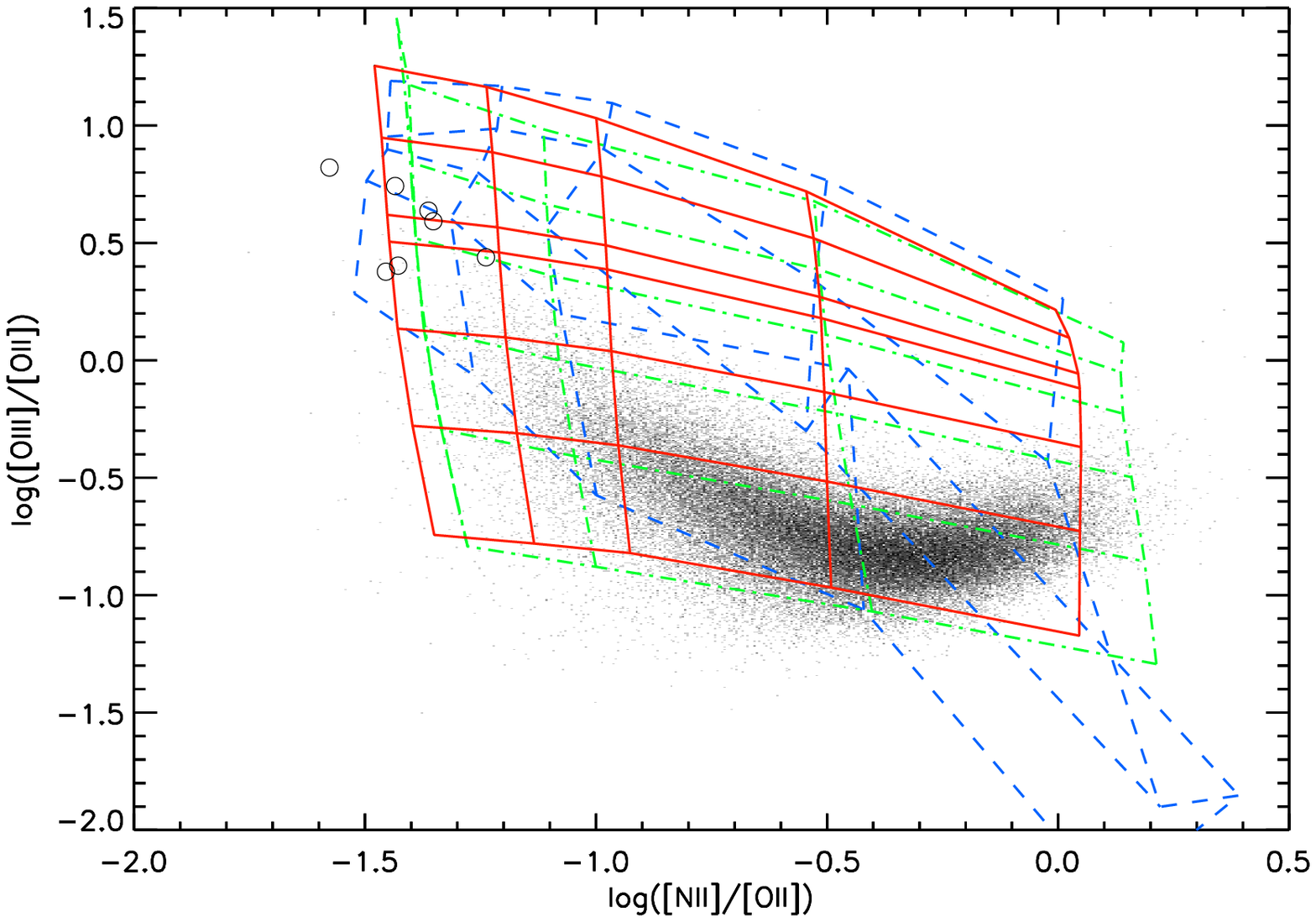}
\includegraphics[width=8.5cm]{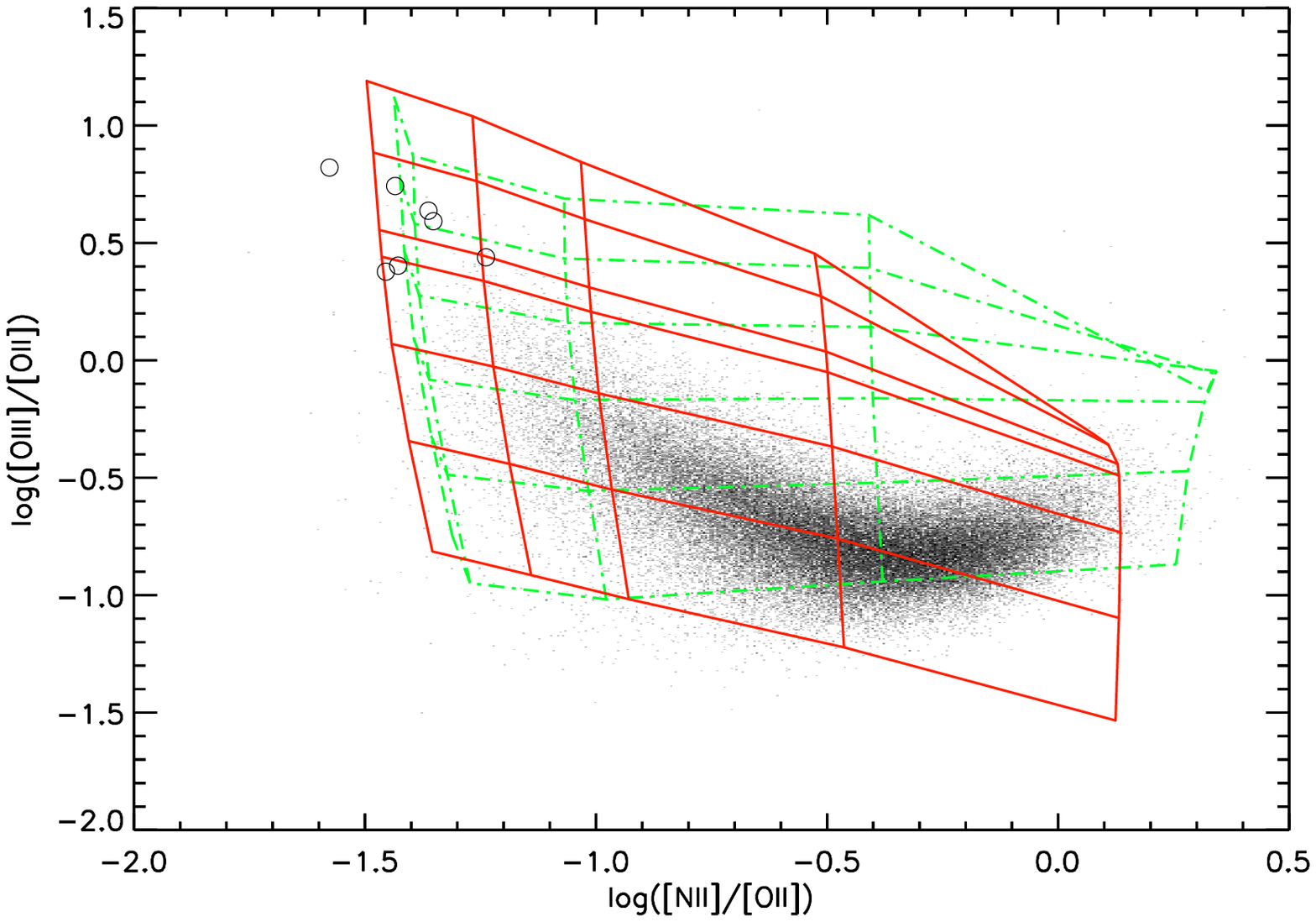}
\includegraphics[width=8.5cm]{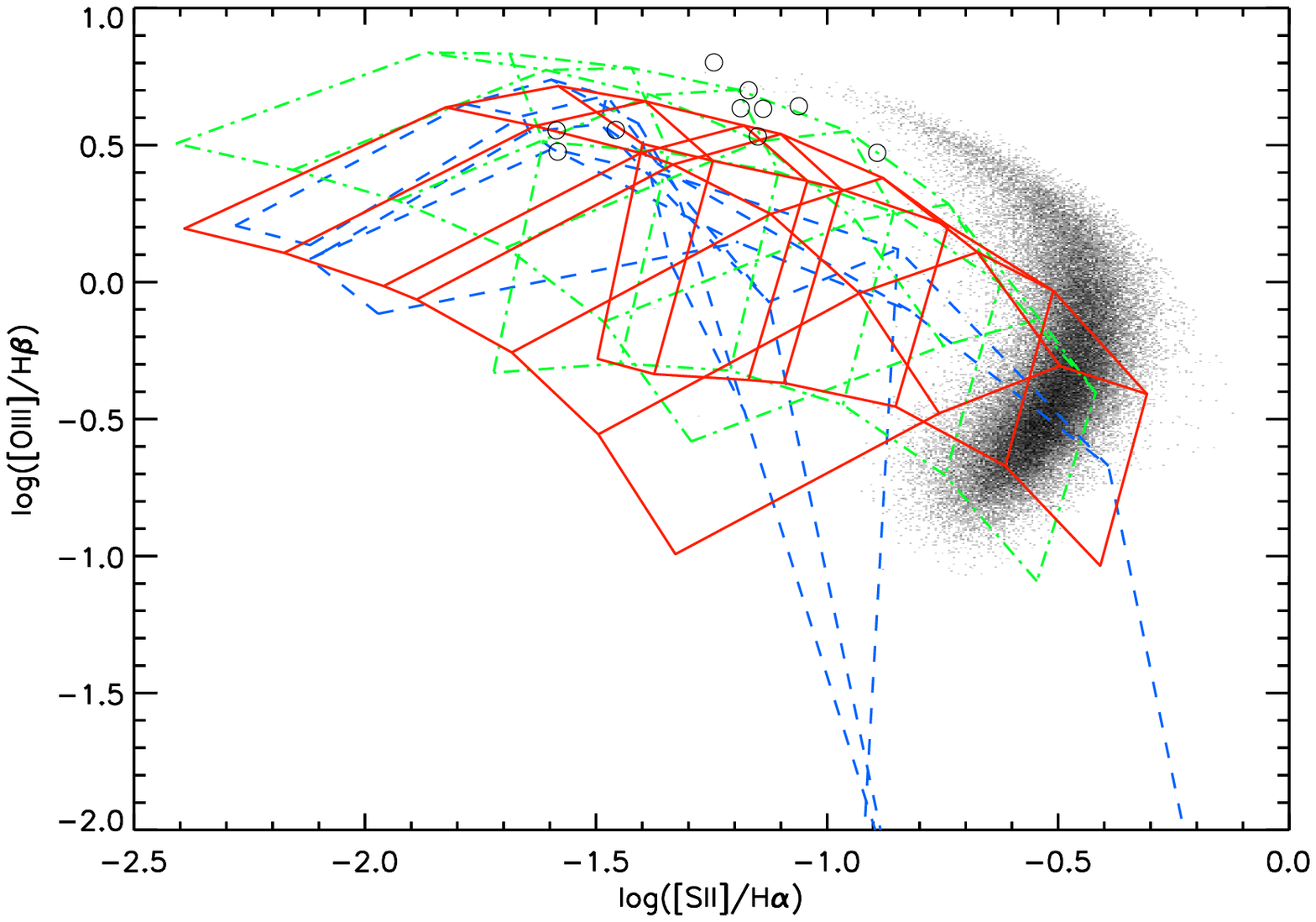}
\includegraphics[width=8.5cm]{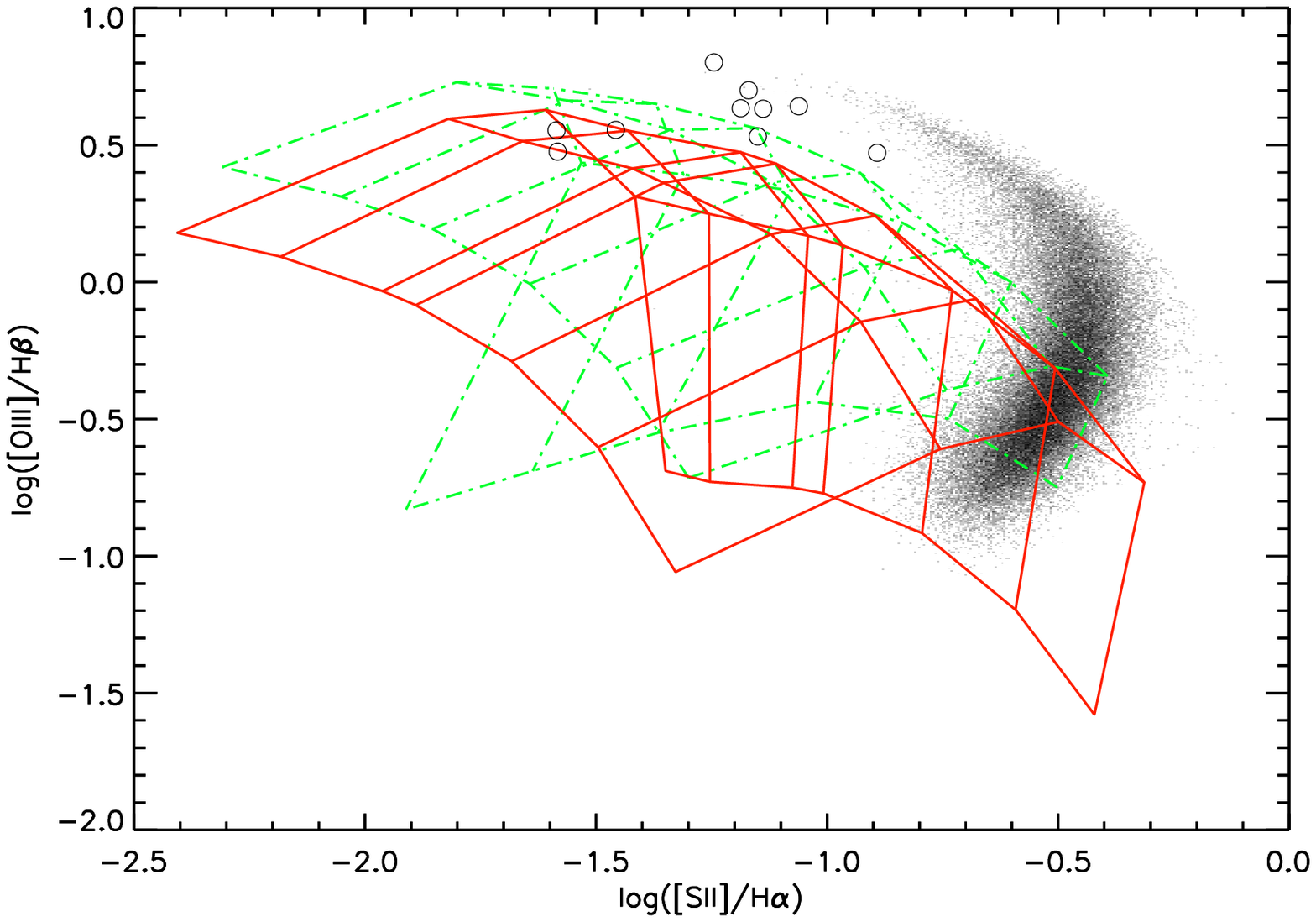}
\caption{{\it Left}: Comparison of instantaneous burst model grids from Kewley et al.\ (2001; green dash-dotted lines), Dopita et al.\ (2006) (blue; dashed lines), and this work (red solid lines) for the [NII]/H$\alpha$ vs. [OIII]/H$\beta$ (top), [NII]/[OII] vs. [OIII]/[OII] (middle), and [SII]/H$\alpha$ vs. [OIII]/H$\beta$ (bottom) diagnostic diagrams. The Kewley et al.\ (2001) diagnostics range from $q = 1 \times 10^7$ to $q = 3 \times 10^8$, and use the Starburst99 stellar population synthesis models and the Geneva evolutionary tracks. The Dopita et al.\ (2006) grids range from 0.0 to 4.0 Myr in increments of 1.0 Myr. The models are plotted with the SDSS galaxies (Kewley et al.\ 2006) and metal-poor galaxies (Brown et al.\ 2006). {\it Right}: Comparing the 8.0 Myr Starburst99/Geneva model grids from Kewley et al.\ (2001) to the 5.0 Myr continuous SFH models from this work (red solid lines). The models are plotted with the SDSS galaxies (Kewley et al.\ 2006) and metal-poor galaxies (Brown et al.\ 2006).}
\end{figure}

\begin{figure}
\centering
\includegraphics[width=15cm]{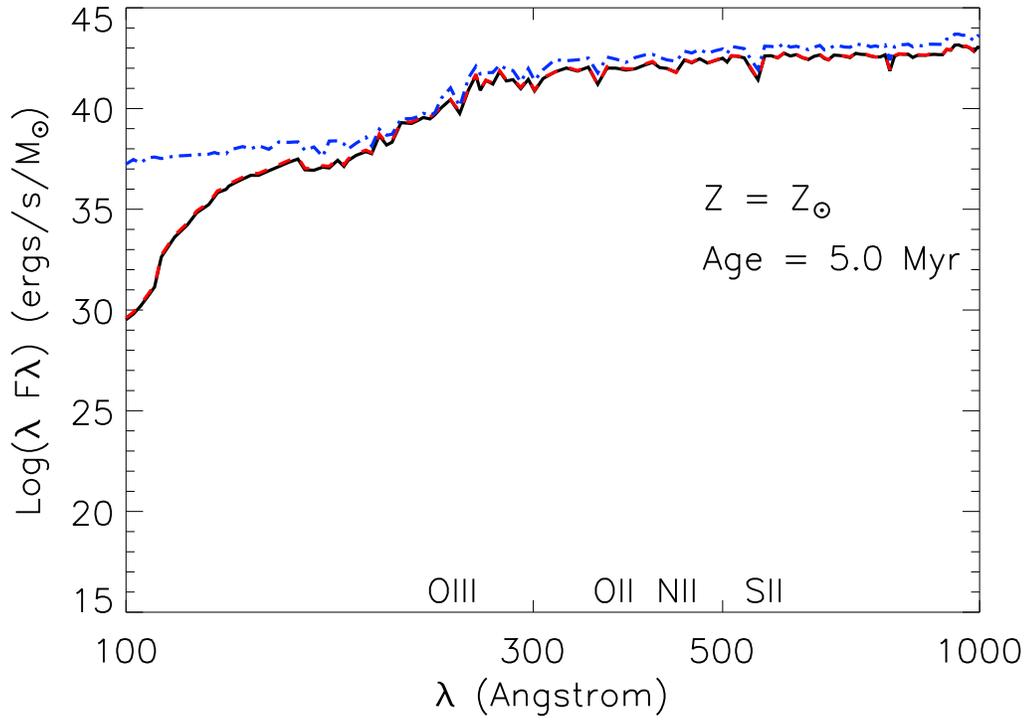}
\caption{Comparison of the FUV ionizing spectrum generating by Starburst99 when adopting the Geneva HIGH tracks (solid black line), the Geneva STD tracks (dashed red line), and the newest generation of the Geneva tracks which include the effects of rotation (dashed-dotted blue line). It is apparent that the rotating tracks generate a much harder ionizing spectrum, particularly in the high-energy regime. All of these tracks are at a metallicity of $Z = Z_{\odot}$. An age of 5.0Myr, a continuous SFH, and $n_e$ = 100 are assumed.}
\end{figure}

\end{document}